\documentclass[english,pra,showpacs,showkeys,tightenlines,secnumarabic,11pt]{revtex4}
\usepackage[T1]{fontenc}
\usepackage[latin1]{inputenc}
\usepackage{amsmath}
\usepackage{graphicx}
\usepackage{amssymb}
\usepackage{epsfig}
\usepackage{graphics}
\usepackage[mathscr]{euscript}
\usepackage{psfrag}
\usepackage{pstricks}
\usepackage{pst-node}
\usepackage{slashed}
\usepackage{babel}

\begin{document}
\title{Double Parton Scatterings in High-Energy Proton-Nucleus Collisions and Partonic Correlations}
\author{S. Salvini, D. Treleani}
\email{simona.salvini@ts.infn.it, daniele.treleani@ts.infn.it} \affiliation{ Dipartimento di
Fisica del{{l'}}Universit\`a di Trieste and INFN, Sezione di
Trieste,\\ Strada Costiera 11, Miramare-Grignano, I-34151 Trieste,
Italy.}
\author{G. Calucci}
\email{giorgio.calucci@ts.infn.it; Retired} \affiliation{Dipartimento di Fisica del{{l'}}Universit\`a, Trieste}

\begin{abstract}
The joint study of Double Parton Scatterings, in high energy proton-proton and proton-nucleus collisions, can provide a lot of information on multi-parton correlations. The multi-parton structure is in fact probed in a different way by DPS, in $p$\,-$p$ and in $p$\,-$A$ collisions. In $p$\,-$A$ collisions the interpretation of the experimental results may be however complicated by the presence of interference terms, which are missing in $p$\,-$p$. A suitable reaction channel, where interference terms are absent, is $WJJ$ production. By studying $WJJ$ production in $p$\,-$Pb$ collisions, we estimate that the fraction of events due to DPS may be larger by a factor 3 or 4, as compared to $p$\,-$p$, while the amount of the increased fraction can give information on the importance of different correlation terms.
\end{abstract}


\maketitle

\section{Introduction}

Multiple Parton Interactions (MPI) have been introduced in the dynamical description of hadronic collisions as a natural solution of the unitarity problem, originated at high energies by the rapid growth of the hard cross sections at small $x$\cite{Sjostrand:1987su, Ametller:1987sg, Ametller:1987ru}. The inclusive cross section is in fact proportional to the multiplicity of elementary partonic interactions and the increasingly large values of the cross section are in this way understood as the result of an increasingly large average number of elementary interactions in an inelastic event. Each elementary partonic interaction is localized in transverse space, inside the much larger overlap region of the matter distribution of the colliding hadrons. A given final state can of course be produced by different MPI processes, which contribute to the cross section with different weights. At small $x$, the leading contribution is provided by the term that maximizes the number of interacting partons, which corresponds to the processes where the hard component of the interaction is maximally disconnected. In the simplest case, namely in Double Parton Scattering (DPS), the dominant contribution to the cross section at small $x$ is thus given by the term where two different pairs of partons interact independently in two different points in transverse space.

As recently pointed out\cite{Blok:2011bu, Blok:2013bpa}, the typical back to back configuration of four large $p_t$ partons, produced in a DPS by the leading contribution at small $x$ and utilized as a distinctive signature for the experimental search of DPS events, can be generated also by a hard interaction involving three partons in the initial state, all localized in the same point in transverse space. The initial partonic flux in $3\to4$ processes is very different as compared to the initial partonic flux of $(2\to2)^2$ processes and, as discussed in \cite{Blok:2011bu}, the typical unbalance of the final state parton pairs is rather different in the two cases. The initial state partonic flux is a measurable quantity and a careful study of the dependence of the cross section on the initial state fractional momenta and on the momentum unbalance should be able to separate experimentally the contributions, to the observed DPS cross section, due to $3\to4$ processes from the leading ones at small $x$. In the present paper we will focus on the disconnected component of DPS in $p$\,-$p$ and in $p$\,-$A$ collisions, while the problem of the experimental identification and subtraction of a possible $3\to4$ background to the observed DPS cross section lies outside our scopes and will not be discussed in the present paper. With DPS we will thus refer specifically to the contribution to the inclusive cross section due to disconnected hard interactions.

When dealing with disconnected hard interactions, the simplest assumption, which leads to very compact results, is that the different partons with small $x$ in the proton are uncorrelated with each other. On the other hand several different types of correlations can be expected and the topic of correlations has addressed a lot of attention\cite{Mekhfi:1985dv, Flensburg:2011kj, Rogers:2009ke, Corke:2011yy, Domdey:2009bg, Diehl:2011tt, Diehl:2011yj, Manohar:2012jr, Chang:2012nw, Rinaldi:2013vpa, Kasemets:2012pr, Diehl:2013mla}. In addition to the dependence on the kinematical variables, DPS amplitudes are in fact expected to depend on spin and color, which induce interference terms in the cross section. Color-correlations are Sudakov suppressed, and thus small for double parton scattering at high energies\cite{Mekhfi:1985dv, Manohar:2012jr}. More important are probably spin correlations, which are expected to affect the rate of double parton scattering and the angular distribution of the final state in particular reaction channels, even if partons are not polarized\cite{Kasemets:2012pr}.

The study of correlations in DPS is therefore a rather rich topic and a proper approach to the problem requires the introduction of flavor and spin dependent double parton distribution functions. The issue is still under theoretical investigation, while present experimental results cannot yet provide indications on the relative importance of different spin and flavor contributions to the double parton distributions. A complete description of DPS in $p$\,-$A$ collisions, even without considering possible $3\to4$ contributions, including however all terms required to account for the dependence on flavor and spin, is therefore still premature. On the other hand there are reasons to expects that, in $p$\,-$A$ collisions, the basic features of DPS will change substantially as compared with the case of DPS in $p$\,-$p$\cite{Strikman:2001gz, Calucci:2010wg, Treleani:2012zi, Blok:2012jr, d'Enterria:2012qx, d'Enterria:2013ck}. It may be therefore instructive to work out explicitly the expectations of DPS in $p$\,-$A$ collisions, when assuming the simplest possible scenario, still not in contradiction with present experimental evidence of DPS in $p$\,-$p$ collisions.

The simplest possibility is to neglect the effects of spin and color in the disconnected component of the DPS cross section. The DPS cross section is thus factorized into functions, which depend on the fractional momenta of the interacting partons, on the resolution of the hard processes and on the relative transverse distance $\beta$ between the two interaction points. The expression of the cross section, for two parton processes $A$ and $B$ in a $p$\,-$p$ collision, is thus given by

\begin{eqnarray}\label{eq:sdouble}
\sigma_{D}^{pp\ (A,B)}&=&\frac{m}{2}\sum_{i,j,k,l}\int\Gamma_{i,j}(x_1,x_2;\beta)\hat\sigma_{i,k}^A(x_1,x_1')\\
&&\qquad\qquad\times\hat\sigma_{j,l}^B(x_2,x_2')\Gamma_{k,l}(x_1',x_2';\beta)\ dx_1dx'_1dx_2dx'_2d^2\beta\nonumber
\end{eqnarray}

where $\Gamma_{i,j}(x_1,x_2;\beta)$ are the double parton distribution functions and the dependence on the fractional momenta of the interacting partons, $x_{1,2}$, and on their relative transverse distance $\beta$ is explicitly indicated, while the dependence on the scales of the two hard processes $A$ and $B$ is understood. The indices $i$ and $j$ label partons flavors. For identical interactions $m=1$ and $m=2$ otherwise. $\hat\sigma^{A}$, $\hat\sigma^{B}$ are the two elementary cross sections.

Eq.(\ref{eq:sdouble}) may lead to a very simple expression, were the cross section is given by the product of the two single scattering inclusive cross sections of the hard processes $A$ and $B$:

\begin{eqnarray}\label{eq:double}
\frac{\sigma_{D}^{pp\ (A,B)}}{dx_idx'_idp_{ti}}=\frac{m}{2}\frac{1}{\sigma_{eff}}\frac{d\sigma^{A}}{dx_1dx'_1dp_{t1}}\frac{d\sigma^{B}}{dx_2dx'_2dp_{t2}}
\end{eqnarray}

which is the "pocket formula" utilized in all experimental analyses of DPS\cite{Akesson:1986iv, Abe:1997xk, Abazov:2009gc, Aad:2013bjm, CMS:awa}. All unknowns in the process converge in this way in the value of a single quantity with the dimensions of a cross section, $\sigma_{eff}$, which is therefore expected to depend on fractional momenta, resolution, parton flavors and on the two-body correlation parameters, which characterize the double parton distributions.

Eq.(\ref{eq:double}) has a transparent physical meaning. When hard interactions are rare, the probability of having also the process $B$ in an inelastic interaction
is given by the ratio $\sigma^B/\sigma_{inel}$. Once the process $A$ takes place, the probability of having the process $B$
in the same inelastic interaction is different. It can anyway be always written as $\sigma^B/\sigma_{eff}$, where
$\sigma_{eff}$ plays {\it effectively} the role, which was of the inelastic cross section in the unbiased case.

Notice that, although $\sigma_{eff}$ is related to the transverse distance between the two hard interactions, it cannot be understood as the effective transverse interaction area, since $\sigma_{eff}$ depends also on the multi-parton distributions in multiplicity. While initial fractional momenta and resolution are measured in the final state and the dependence of $\sigma_{eff}$ on parton flavors can be obtained, at least to a certain extent, by selecting different reaction channels, the effects of the dependence on the partonic distributions in multiplicity and on the correlation in the relative transverse distance, cannot be disentangled by looking only at $p$\,-$p$ collisions.

Even in the simplest scenario, $\sigma_{eff}$ is thus expected to depend on flavor and on all kinematical variables. In spite of that, Eq.(\ref{eq:double}) has shown to be able to describe the experimental results of the direct search of double parton collisions in rather different reaction channels and kinematical regimes\cite{Akesson:1986iv, Abe:1997xk, Abazov:2009gc, Aad:2013bjm, CMS:awa} with a value of $\sigma_{eff}$ not incompatible with a universal constant, while the study of CDF\cite{Abe:1997xk}, of the dependence of $\sigma_{eff}$ on the fractional momenta of the incoming partons, is again not inconsistent with a value of $\sigma_{eff}$ independent on $x$. When the DPS cross section is generalized by introducing parton distributions depending on transverse momenta and off shell T-matrix elements, the same value of $\sigma_{eff}$ allows describing DPS also in the regime of very small $x$, where the back to back kinematical configuration, typical of the large $p_t$ partons originated by DPS, is lost. The observed production rates of $({\rm J}/\psi,\ {\rm J}/\psi)$ are thus understood\cite{Baranov:2011ch}, while the production of different combinations of charmed mesons, the differential distributions in the $D^0D^0$ invariant mass and the azimuthal correlation between two $D^0$ mesons, as worked out in\cite{Maciula:2013kd}, are not incompatible with the recent measurements of the LHCb Collaboration\cite{Aaij:2012dz}.

One should underline that the experimental indication, of a value of $\sigma_{eff}$ consistent with a universal constant, represents a non trivial test of the simple interaction mechanism leading to Eq.\eqref{eq:double}. The expression of  $\sigma_{D}$ in Eq.\eqref{eq:double} depends in fact rather strongly on the kinematical conditions of the observed process. In particular the dependence on the incoming parton flux is much stronger as compared with the case of a single hard scattering process.

As already noticed, although $\sigma_{eff}$ is directly related to parton correlations, even in the simplest scenario, by measuring DPS only in $p$\,-$p$ collisions one does not have enough information to decide how much the observed value of $\sigma_{eff}$ is originated by the typical separation in transverse space between the two pairs of interacting partons and how much it is rather due to the actual distribution in multiplicity of parton pairs in the hadronic structure. Additional information to discriminate between the two cases can be nevertheless obtained by studying DPS in $p$\,-$A$ collisions. MPI in $p$\,-$A$ collisions introduce in fact novel features in the process. A relevant novel feature is that one may have MPI, where two or more target nucleons are active participants in the hard process\cite{Strikman:2001gz, Calucci:2010wg, Treleani:2012zi, Blok:2012jr, d'Enterria:2013ck}. A relevant consequence of having two or more active target nucleons is that while in the simplest picture of the interaction considered here, in $p$\,-$p$ collisions MPI are described by the incoherent superposition of sets of elementary partonic interactions\cite{Sjostrand:1987su, Ametller:1987sg, Ametller:1987ru}\cite{Calucci:2009ea}, in $p$\,-$A$ collisions interference terms may, on the contrary, play an important role\cite{Treleani:2012zi}.

To have some quantitative indication on the impact of the different features of DPS in $p$\,-$A$ collisions, we will study the simplest option, where in $p$\,-$p$ collisions $\sigma_{eff}$ is a universal constant and it is completely determined by the typical transverse distance, between the two pairs of interacting partons, and by the multiplicity of parton pairs in the hadronic structure. We will further simplify the problem by selecting a suitable reaction channel, where there are no contributions of interference terms in $p$\,-$A$ collisions. One can then show that the multiplicity of pairs of partons and their typical transverse separation have rather different effects on the DPS cross section in $p$\,-$p$ and in $p$\,-$A$ collisions. The amount of increase of the cross section when going to $p$\,-$A$ can be in fact linked in a rather direct way to the multiplicity of parton pairs in the projectile, while the effects of the typical separation of the parton pairs in transverse space are only of minor importance.

The paper is organized as follows. In the next section we recall some of the main features of DPS in $p$\,-$p$ and in $p$\,-$A$ collisions. In the following section we discuss the case of $WJJ$ production. A section is devoted to illustrate, with some numerical estimates, the different effects on the $p$\,-$A$ cross section of varying either the multiplicity of pairs of partons or their relative transverse distance. The last section is dedicated to the summarizing remarks.

\section{DPS in $p$\,-$p$ and in $p$\,-$A$ collisions}

It has been pointed out that, for sufficiently small values of $\beta$, the Distributions $\Gamma(x_1,x_2;\beta)$ can be expressed in terms of known quantities\cite{Diehl:2011yj}. For small $\beta$, $\Gamma(x_1,x_2;\beta)$ may be obtained from a single-parton distribution times a perturbative dynamics, which yields the splitting function for the longitudinal variables and a $1/\beta^2$ singularity in the transverse relative distance. The divergent behavior of the DPS cross section at small $\beta$ needs therefore to be properly subtracted and the subtraction terms included in the single scattering contribution. The issue of subtraction of the divergent contribution and of the correlation in fractional momenta, induced by perturbative splitting, has been discussed by several authors and is still a matter of debate, in particular for what concerns the QCD evolution of the double parton distribution functions\cite{Blok:2011bu, Blok:2013bpa,Diehl:2011yj,Gaunt:2011xd,Ryskin:2012qx,Ryskin:2011kk}.

The common origin of the initial state partons, in the small $\beta$ region, leaves anyway track in the DPS cross section.
A main qualitative feature is the presence of additional contributions, which however cannot be considered any more as disconnected in transverse space and cannot be expressed by Eq.(\ref{eq:sdouble}). The importance of these contributions grows with the fractional momenta of the incoming partons and would thus induce a measurable dependence of $\sigma_{eff}$ on the initial state fractional momenta. Although there was no systematic study of the $x$ dependence, the available experimental evidence does not seem to imply a sizable dependence of $\sigma_{eff}$ on the initial state fractional momenta. As stated in the introduction, we will therefore take the simplified attitude of assuming that possible additional contributions to the measured DPS cross section can be identified and subtracted experimentally and we will focus on the disconnected DPS interaction mechanism.

Disregarding for simplicity the dependence on flavor and on the resolution, one may introduce

\begin{eqnarray}\label{eq:kappa}
G(x_1,x_2)\equiv\int\Gamma(x_1,x_2;\beta)d^2\beta,\qquad G(x_1,x_2)\equiv K_{x_1x_2}G(x_1)G(x_2)
\end{eqnarray}

where $G(x)$ are the usual one-body distribution functions. Without any loss of generality, one may thus write

\begin{eqnarray}\label{eq:gamma2}
\Gamma(x_1,x_2;\beta)=K_{x_1x_2}G(x_1)G(x_2)f_{x_1x_2}(\beta)
\end{eqnarray}

with $\int f_{x_1x_2}(\beta)d^2\beta=1$. One has

\begin{eqnarray}
\sigma_{D}^{pp\ (A,B)}(x_1,x'_1,x_2,x'_2)&=&\frac{m}{2}K_{x_1x_2}K_{x'_1x'_2}G(x_1)\hat\sigma_A(x_1,x'_1)G(x'_1)\cr
&&\qquad\times G(x_2)\hat\sigma_B(x_2,x'_2)G(x'_2)
\int f_{x_1x_2}(\beta)f_{x'_1x'_2}(\beta)d^2\beta\cr
&=&\frac{m}{2}\frac{K_{x_1x_2}K_{x'_1x'_2}}{\pi\Lambda^2(x_1,x'_1,x_2,x'_2)}\sigma_A(x_1,x'_1)\sigma_B(x_2,x'_2)
\nonumber
\end{eqnarray}

where

\begin{eqnarray}\label{eq:Lambda}
\int f_{x_1x_2}(\beta)f_{x'_1x'_2}(\beta)d^2\beta=\frac{1}{\pi\Lambda^2(x_1,x'_1,x_2,x'_2)}
\end{eqnarray}

The effective cross section is therefore given by

\begin{eqnarray}\label{eq:seff}
\sigma_{eff}(x_1,x'_1,x_2,x'_2)=\frac{\pi\Lambda^2(x_1,x'_1,x_2,x'_2)}{K_{x_1x_2}K_{x'_1x'_2}}
\end{eqnarray}

and $\Lambda(x_1,x'_1,x_2,x'_2)$ measures the typical transverse distance between the pairs of interacting partons, for given values of fractional momenta, while $K_{x_1x_2}$ gives the second moment of the multiparton exclusive multiplicity distribution. More precisely $K_{x_1x_2}=\langle n(n-1)\rangle_{{x_1}{x_2}}/\bigl(\langle n\rangle_{x_1}\langle n\rangle_{x_2}\bigr)$\cite{Calucci:2010wg} in such a way that, in the simplest case of a Poissonian distribution in multiplicity, one would have $K_{x_1x_2}=1$. Present experimental indication is that the effective cross section depends only weakly on fractional momenta.

As apparent in Eq.(\ref{eq:seff}), in nucleon-nucleon collisions all effects due to parton correlations are summarized in the value of a single quantity (the effective cross section) and nucleon-nucleon collisions alone do not allow to measure $\Lambda$ and $K$ separately. To obtain additional information on multi-parton correlations one needs to study DPS $p$\,-$A$ collisions.

Obviously Double (and more in general Multiple) Parton Scatterings are more abundant in reactions with nuclei. DPS are thus more interesting in $p$\,-$A$ collisions. When non additive corrections to the nuclear parton distributions are only a minor effect, in $p$\,-$A$ collisions DPS originate either from interactions with a single active target nucleon or from interactions with two different active target nucleons. While the first contribution does not add much to the information already available from DPS on a isolated nucleon, the second contribution has the peculiar property of {\it enhancing the effects of longitudinal correlations in the proton}. In the latter case the relative transverse distance between the interacting pairs does not play in fact any relevant role, when compared to the much larger nuclear radius\cite{Strikman:2001gz, Calucci:2010wg, Treleani:2012zi, Blok:2012jr, d'Enterria:2012qx, d'Enterria:2013ck}. By selecting the contribution to DPS, with two active target nucleons, one will hence have direct access to the longitudinal correlations of the hadron structure.
The cross section thus splits into two terms:

\begin{eqnarray}\label{eq:DpA}
\sigma_{D}^{pA}=\sigma_{D}^{pA}\big |_1+\sigma_{D}^{pA}\big |_2
\end{eqnarray}

which correspond to the two different contributions, where the double hard interaction takes place with one or with two different target nucleons. In a simplest probabilistic picture of the interaction one would write

\begin{eqnarray}\label{eq:Glauber}
\sigma_{D}^{pA}\big |_1=\frac{1}{2}\frac{\sigma_S^2}{\sigma_{eff}}\int d^2B\ T(B) \propto A,\qquad\sigma_{D}^{pA}\big |_2=\frac{1}{2}\sigma_S^2\int d^2B\ T^2(B)\propto A^{4/3}
\end{eqnarray}

\noindent
where the case of two identical partonic interactions has been considered. Here $\sigma_S$ is the inclusive single scattering cross section and $T(B)$ is the nuclear thickness, as a function of the impact parameter of the collisions $B$. The two terms have a transparent geometrical meaning and are distinguished by their different dependence on the atomic mass number $A$.

\begin{figure}[h]
\centering
\includegraphics[width=140mm]{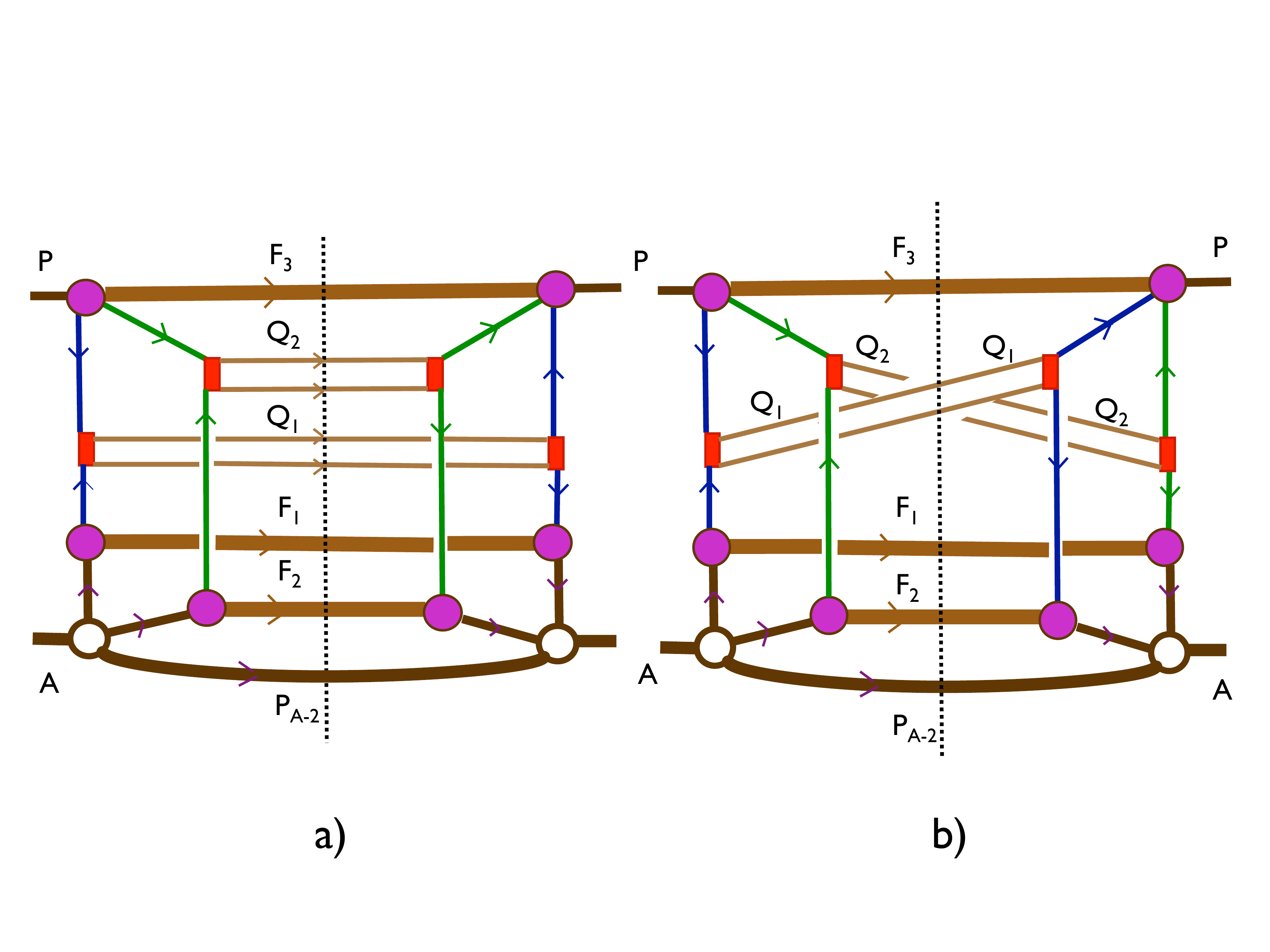}
\caption {Double parton scattering contributions to the discontinuity of the forward $p$\,-$A$ interaction amplitude: a) diagonal term, b) interference term}
\label{fig:diagram_a}
\end{figure}

A closer look at the kinematics of the process\cite{Treleani:2012zi, Blok:2012jr}, shows however that one needs to take into account an additional contribution to the cross section. The diagonal term $\sigma_{D}^{pA}\big |_2$ in Eq.(\ref{eq:Glauber}), does not exhaust in fact all possibilities of interaction and, in the case of two different active target nucleons, one needs to add an interference term. The two terms are conveniently expressed as contributions to the discontinuity of the forward elastic amplitude. The two corresponding unitarity diagrams are shown in Fig.\ref{fig:diagram_a}, which represents the diagonal, Fig.\ref{fig:diagram_a} a), and the off-diagonal contribution, Fig.\ref{fig:diagram_a} b) \footnote{Notice that having focused on the disconnected component of the hard interaction, the description of the process is greatly simplified, with respect to the sizably more complex situation discussed in\cite{Blok:2012jr}.}.

Following the lines described in full detail in \cite{Calucci:2010wg, Treleani:2012zi}, the contribution of the diagonal term to the cross section, with two active target nucleons, is given by:

\begin{eqnarray}\label{eq:corr}
\sigma^{pA}_{D}\big |_{2,\, diag}&=&\frac{1}{2(2\pi)^3}\int\Gamma(x_1,x_2;\beta_1-\beta_2)
\frac{d\hat\sigma(x_1, x_1')}{d\Omega_1}
\frac{d\hat\sigma(x_2, x_2')}{d\Omega_2}
\Gamma( x_1'/ Z_1;b_1)\Gamma(x_2'/Z_2;b_2)\cr
&\times &|\tilde\Psi_A( Z_i;B_i)|^2 db_1\, db_2\,d(\beta_1-\beta_2)\delta(B_1-B_2+b_1 -b_2+\beta_1-\beta_2)\cr
&\times &\delta \Bigl(\sum Z_i-A\Bigr)\,dx_1dx_2dx_1'dx_2'\,d\Omega_1\,d\Omega_2\prod_i dB_i\frac{d Z_i}{Z_i}
\end{eqnarray}

\noindent
where $\hat\sigma(x_i, x_i')$ are the partonic cross sections and the nuclear wave function $\Psi_A( Z_i;B_i)$ is in a mixed representation, with  $Z_i$ the nucleon's fractional momenta and $B_i$ the nucleon's transverse coordinates. The nuclear wave function is peaked at $Z_i=1$, while MPI are most important at $x\simeq10^{-2}\div10^{-3}$. Keeping into account that the scale of the nucleon's Fermi momentum is small as compared to the nucleon mass, a meaningful approximation is to integrate on $Z_i$ while keeping $Z_1=Z_2=1$ in the partonic distributions $\Gamma$. The nuclear dependence is thus expressed through the two-body nuclear density $\rho(B_1, z_1;\, B_2,z_2)$, where the quantities $z_1$ and $z_2$ are the longitudinal coordinates of the two interacting nucleons:

\begin{eqnarray}\label{eq:corr1}
&&\int\Gamma( x_1'/ Z_1;b_1)\Gamma(x_2'/Z_2;b_2)|\tilde\Psi_A( Z_i;B_i)|^2 \prod_i \frac{d Z_i}{Z_i}\cr
&&\qquad\qquad\simeq\Gamma( x_1';b_1)\Gamma(x_2';b_2)\, \int \rho(B_1, z_1;\, B_2,z_2)dz_1dz_2
\end{eqnarray}

\begin{figure}[h]
\centering
\includegraphics[width=130mm]{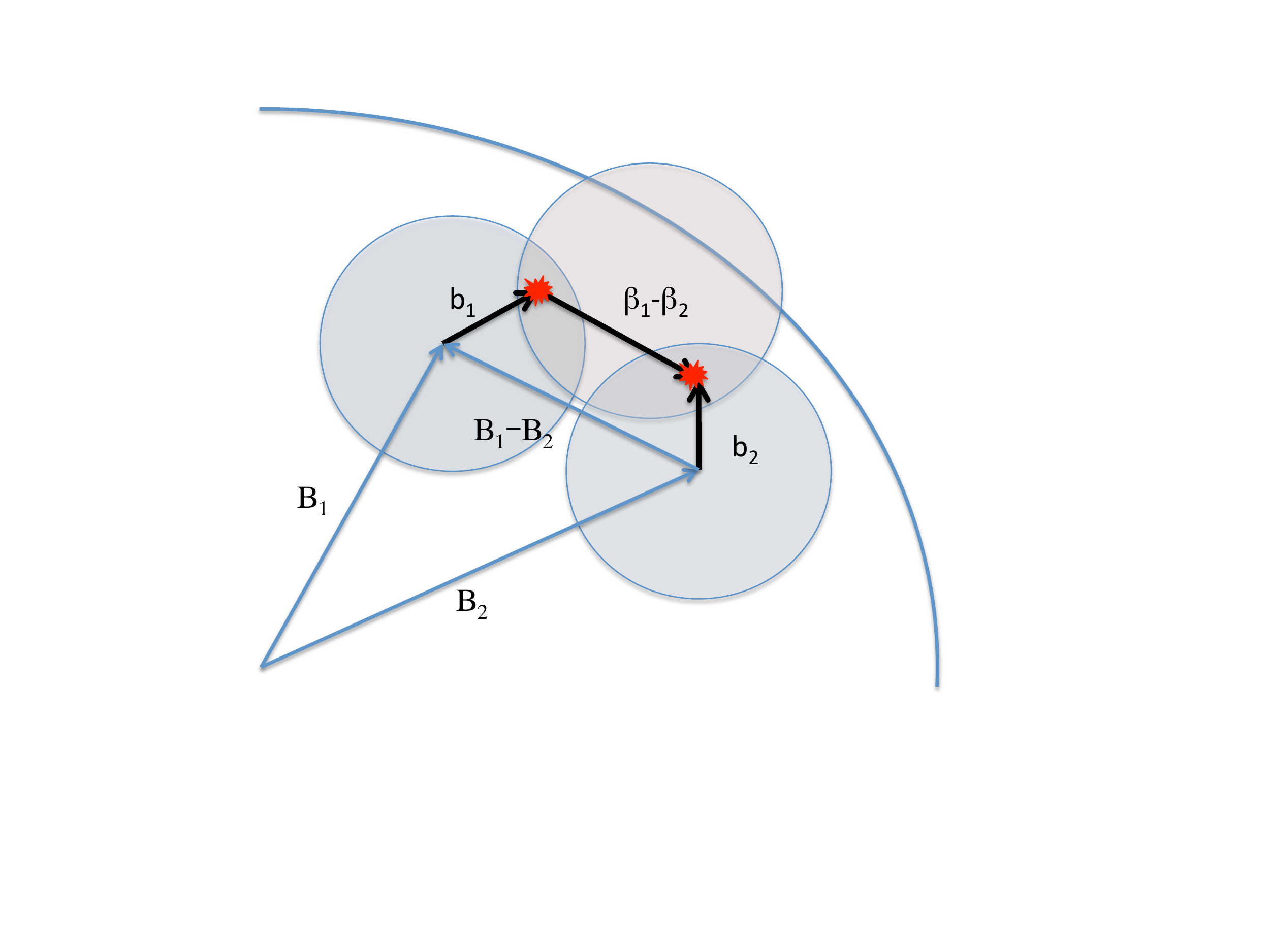}
\caption {Configuration in transverse space corresponding to the DPS cross section in Eq.(\ref{eq:correlation})}
\label{fig:correlation}
\end{figure}

With the help of Eq.(\ref{eq:gamma2}), in the case of two identical interactions, one thus obtains

\begin{eqnarray}\label{eq:correlation}
\frac{d\sigma^{pA}_{D}\big |_{2,\, diag}}{dx_idx_i'd\Omega_i}&=&K_{x_1x_2}\frac{1}{2}
\frac{d\sigma_S(x_1, x_1')}{d\Omega_1}
\frac{d\sigma_S(x_2, x_2')}{d\Omega_2}
\int f_{x_1x_2}(\beta_1-\beta_2)f_{x_1'}(b_1)f_{x_2'}(b_2)\cr
&\times & \rho(B_1, z_1;\, B_2,z_2)\, dz_1dz_2\, \delta(B_1-B_2+b_1 -b_2+\beta_1-\beta_2)\cr
&\times & db_1\, db_2\,d(\beta_1-\beta_2)\,dB_1\,dB_2
\end{eqnarray}

\noindent
where $\sigma_S$ is the usual single scattering inclusive cross section on a nucleon and we made the positions $\Gamma( x;\,b)\equiv G(x)f_x(b),\quad\int f_x(b)d^2b=1$. The configuration in transverse space corresponding to the DPS cross section in Eq.(\ref{eq:correlation}) is illustrated in Fig.\ref{fig:correlation}.

The contribution of the interference term to the cross section is obtained in a similar way\cite{Treleani:2012zi}. The expression is

\begin{eqnarray}\label{eq:offdiag}
\frac{\sigma^{pA}_{D}\big |_{2,\, int}}{dx_idx_i'd\Omega_i}&=&\frac{1}{(2\pi)^3}\int\Gamma(x_1,x_2;\beta_1-\beta_2)\frac{d\hat\sigma(x_1, x_1')}{d\Omega_1}
\frac{d\hat\sigma(x_2, x_2')}{d\Omega_2}\cr
&\times&W(Z_1,Z_2; Z_1', Z_2'; x_1', x_2';b_1,b_2;B_1,B_2)\cr
& \times&\tilde\Psi_A( Z_i;B_i) \tilde\Psi_A^*( Z_i';B_i)
\delta(B_1-B_2-b_1 +b_2-\beta_1+\beta_2)\delta(Z_1- Z_1'- x_1'+ x_2')\cr
&\times&\delta(Z_2- Z_2'+ x_1'- x_2')\delta\Bigl(\sum Z_i-A\Bigr)\delta\Bigl(\sum Z_i'-A\Bigr)\cr
&\times&db_1\, db_2\,d(\beta_1-\beta_2)\,\prod dB_i\frac {dZ_i}{Z_i}\frac{d Z_i'}{ Z_i'}
\end{eqnarray}

\noindent
where the off diagonal parton amplitudes in the process are all included in the function $W$. Some detail on the construction of the function $W$ are presented in Appendix A.

\begin{figure}[h]
\centering
\includegraphics[width=130mm]{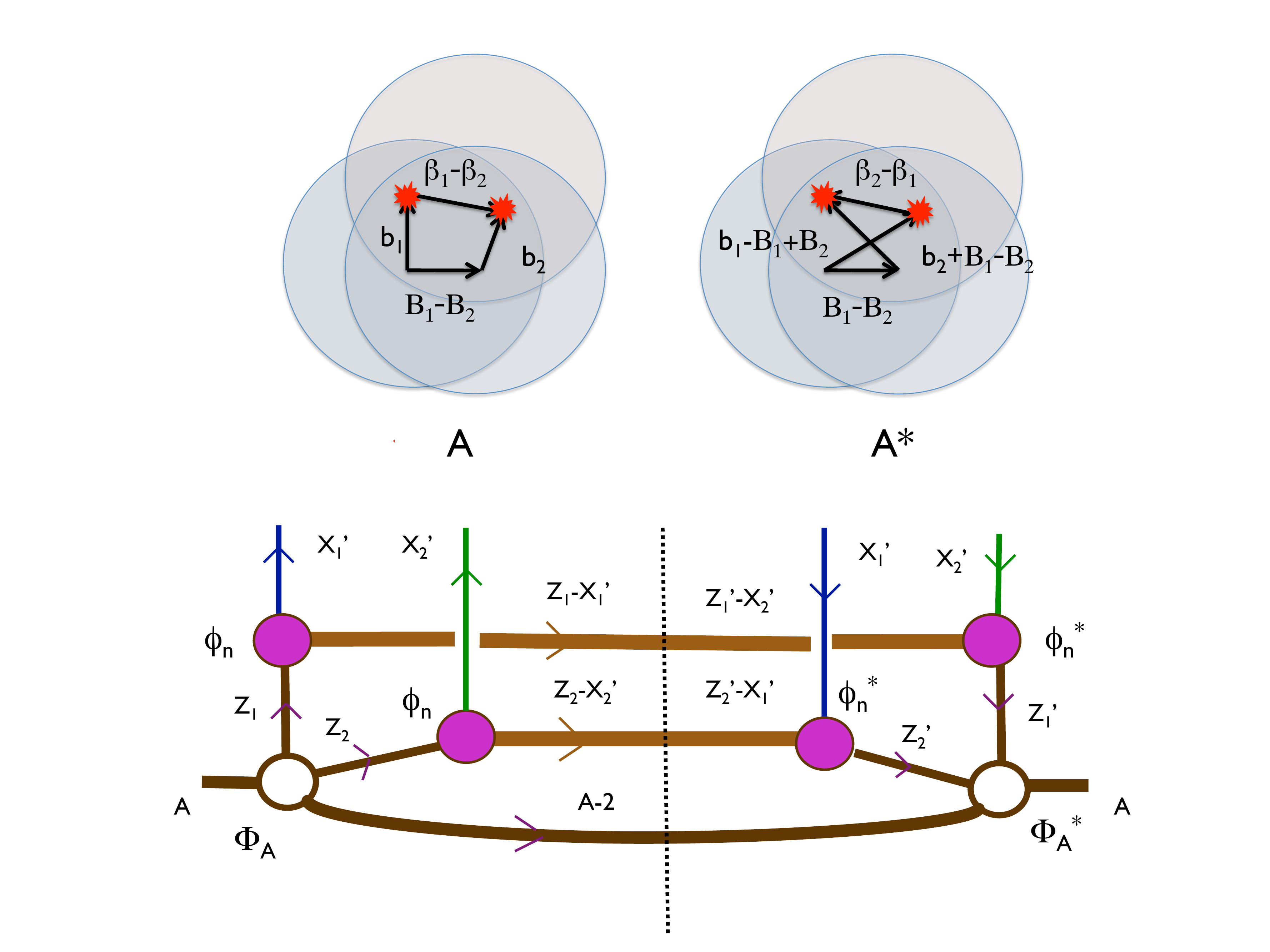}
\caption {Upper part of the figure: configurations in transverse space of the left (A) and right (A*) hand site amplitudes in the off diagonal contribution to the cross section in Eq.(\ref{eq:offdiag}).
Lower part of the figure: nucleon's fractional momentum flow in the off diagonal contribution to the cross section in Eq.(\ref{eq:offdiag}).}
\label{fig:overlap}
\end{figure}

The main features of the off diagonal contribution originate from kinematics and are summarized in Fig.\ref{fig:overlap}.

As far as the longitudinal variables are concerned, the interference term requires the nuclear wave function to be taken at different values of $Z$. As apparent in the lower part of Fig.\ref{fig:overlap}, one must have in fact $Z_1- Z_1'= x_1'- x_2'= Z_2'-Z_2$. When the differences $Z_i- Z_i'$ are not too small, the interference term is therefore depressed with respect to the diagonal term by the nuclear form factor. One should however keep in mind that at large energies the values of $ x$ can be rather small, still maintaining the process within the limits of perturbative  dynamics, so the depression factor may not be strong.

When looking at transverse momenta, there are two different scales in the initial state: The typical transverse momentum of the order of a few GeV of the initial state partons, originated by QCD evolution, and the typical transverse momentum of the order of a hundred of MeV of the bound nucleons, due to Fermi motion. One might therefore expect the interference term to be further depressed by the presence of the two different scales. As discussed in Appendix A, closer look at kinematics shows however that this is not the case. The difference between the overall transverse momenta of the large $p_t$ partons, generated in the two hard collisions, actually ${Q}_{1,t}-{Q}_{2,t}$ in Fig.\ref{fig:Appinterf} in Appendix A, is in fact originated in the upper vertex $\phi_p$ in the figure and does not propagate to the target nucleon's momenta in lower part of the diagram, in such a way that there are not transverse momenta of the GeV scale in the nucleon's lines in the interference diagram. Going to the coordinates space, one obtains a rather transparent picture of the interaction, as shown in the upper part of Fig.\ref{fig:overlap}, where the configuration in transverse space, of the amplitude in the left hand side of the cut of Fig.\ref{fig:diagram_a} b), is labelled with A and the configuration of the amplitude, in the right hand side of the cut, is labelled with A*. By exchanging ${Q}_{1,t}$ with ${Q}_{2,t}$, when moving form the left to the right hand side of the cut in Fig.\ref{fig:diagram_a} b), one produces a change in the sign of the difference between the transverse coordinates of the two projectile partons, $\beta_1$ and $\beta_2$, in the non perturbative vertex $\phi_p$.  The argument of $\phi_p$ is therefore $(\beta_1-\beta_2)$ in the amplitude A and  $(-\beta_1+\beta_2)$ in the amplitude A*, while one does not expect any dependence of $\phi$ on the sign of its argument. For what concerns the remaining transverse variables, as illustrated in the upper part of Fig.\ref{fig:overlap}, the nuclear configurations in A and in A* are the same, the two hard interactions remain localized in the same points and the two interacting partons exchange their parent hadrons.

When the two active partons are identical, the initial partonic configuration of the DPS process can be therefore produced in two independent ways, since each of the two active nucleons can generate each of the two interacting partons. The two nuclear configurations have therefore to be added coherently in the cross section. The two active target nucleons have different longitudinal fractional momenta in the two configurations and the difference is equal to the difference of the fractional momenta of the two interacting partons $x_1'- x_2'$. The interference term is thus characterized by the peculiar dependence of the nuclear form factor as a function of $x_1'- x_2'$. Notice that the interference term is directly proportional to the off diagonal parton distributions, while the value of $x_1'- x_2'$ can be controlled in the process by selecting the kinematical configuration of the final state produced by the hard interaction. There is thus the interesting possibility of obtaining additional information on off diagonal parton distribution functions by using an inclusive process, actually DPS in $p$\,-$A$ collisions, which, comparing with exclusive processes, would be able to provide larger rates of events and to access different kinematical regimes.

   \section {$WJJ$ production by DPS in $p$\,-$A$ collisions}

The presence of an interference term in DPS in $p$\,-$A$ collisions may have an interesting potential to study the generalized parton distributions. On the other hand our present purpose is to identify and work out a simple case, where the measured cross section has a straightforward relation with partonic correlations. We will thus consider a reaction channel where the interference term is strongly suppressed. To our aim a particularly interesting channel is the inclusive production of $WJJ$. In $WJJ$ production, the two initial state partons, both from the side of the projectile and from the side of the target nucleus, are a quark and a gluon since, in the kinematical regime of interest for DPS, $JJ$ production is dominated by the gluonic channel. The two active target partons cannot therefore be identical and the interference term is absent. A further reason of interest in $WJJ$ is that DPS production is presently studied experimentally in $p$\,-$p$ both by ATLAS and by CMS \cite{Aad:2013bjm, CMS:awa} while, after the recent experimental results and in view of the next runs planned at the LHC, there is an increasing activity in $p$\,-$A$ collision both experimental and theoretical\cite{Albacete:2013ei}. The experimental study of DPS in $WJJ$ production in $p$\,-$Pb$ collisions might therefore represent a feasible option for the experimental groups in a not too distant future.

In the inclusive cross section for $WJJ$ production, $\sigma^{pA}(WJJ)$, one identifies three different contributions:

\begin{eqnarray}
\sigma^{pA}(WJJ)=\sigma_{S}^{pA}(WJJ)+\sigma_{D}^{pA}(WJJ),\, {\rm where}\quad \sigma_{D}^{pA}(WJJ)=\sigma_{D}^{pA}(WJJ)\big |_1+\sigma_{D}^{pA}(WJJ)\big |_2
\nonumber
\end{eqnarray}

\noindent
The first term, $\sigma_{S}^{pA}(WJJ)$, represents the processes where $WJJ$ is produced by a single parton collision while the contribution due to DPS, $\sigma_{D}^{pA}(WJJ)$, is expressed, according with Eq.(\ref{eq:DpA}), by the sum of two terms, which distinguish whether the DPS takes place against a single, $\sigma_{D}^{pA}(WJJ)\big |_1$, or against two different target nucleons, $\sigma_{D}^{pA}(WJJ)\big |_2$.

The contribution due to a single parton collisions, $\sigma_{S}^{pA}(WJJ)$, can be evaluated according to the standard rules, keeping into account the different contributions due to the interaction with a target proton or neutron and making use of the parton distributions of the bound nucleons\cite{Eskola:2009uj, Eskola:1998df}. This term does not provide much additional information on hadron structure and it can be considered as a known quantity.

The explicit expression of the contribution due to a double parton scattering in a collision with a single target nucleon, $\sigma_{D}^{pA}(WJJ)\big |_1$, is

\begin{eqnarray}\label{eq:DpA1}
\sigma_{D}^{pA}(WJJ)\big |_1=\frac{1}{\sigma_{eff}}\bigl[Z\sigma_S^{p[p]}(W)\sigma_S^{p[p]}(JJ)+(A-Z)\sigma_S^{p[n]}(W)\sigma_S^{p[n]}(JJ)\bigr]
\end{eqnarray}

where $\sigma_S^{p[p],p[n]}(W)$ are the single scattering cross sections for inclusive production of a $W$ in a collision of a proton with a bound proton or with a bound neutron, while $\sigma_S^{p[p],p[n]}(JJ)$ is, analogously, the single scattering cross section to produce a pair of jets. The effective cross section, $\sigma_{eff}$, has been assumed to be a universal constant. $A$ is the atomic mass number and $Z$ the nuclear charge. Analogously to the term due to a single parton collisions, $\sigma_{S}^{pA}(WJJ)$, also $\sigma_{D}^{pA}(WJJ)\big |_1$ is therefore expressed fully explicitly in terms of known quantities and is evaluated with the standard rules of of the QCD-parton model, with the help of the parton distributions of the bound nucleons. The contribution due to a double parton scattering, in hard collisions with a single target nucleon, does not have much to add to the information on the hadron structure already available from double parton interactions in proton-proton collisions and also this term can thus be regarded as a known contributions to the cross section.

All novel information on hadron structure, provided by DPS in $p$\,-$A$ collisions, is to be found in the last term, $\sigma_{D}^{pA}(WJJ)\big |_2$, where two different nucleons participate to the double parton interaction. According with the discussion in the previous section, the corresponding contribution to the cross section is

\begin{eqnarray}\label{eq:WJJ}
\sigma^{pA}_{D}(WJJ)\big |_{2}&=&K_{x_1x_2}
\sigma_S(W)
\sigma_S(JJ)
\int f_{x_1x_2}(\beta_1-\beta_2)f_{x_1'}(b_1)f_{x_2'}(b_2)\cr
&\times & \rho(B_1, z_1;\, B_2,z_2)\, dz_1dz_2\, \delta(B_1-B_2+b_1 -b_2+\beta_1-\beta_2)\cr
&\times & db_1\, db_2\,d(\beta_1-\beta_2)\,dB_1\,dB_2
\end{eqnarray}

The expression in Eq.(\ref{eq:WJJ}) has been obtained by disregarding the dependence of $\Gamma(x_i'/Z_i;b_i)$ on $Z_i$ in Eq.(\ref{eq:corr1}). The lower limit of the integration on $Z_i$ in Eq.(\ref{eq:corr1}) is $x_i'$ and one has therefore implicitly assumed that $x_i'\ll Z_i$, which limits the validity of Eq.(\ref{eq:WJJ}) to the region of small $x_i'$.

The produced spectrum is directly proportional to the overlap integral in the transverse coordinates. The corresponding configuration is shown in Fig.\ref{fig:correlation}. The overlap integral depends on the three different transverse scales, which characterize $f_{x_i'}(b_i)$, $f_{x_1x_2}(\beta_1-\beta_2)$ and $\int \rho(B_1, z_1;\, B_2,z_2)dz_1dz_2$. When comparing hadronic and nuclear scales, a sensible approximation is to neglect the hadronic scale when compared to the nuclear scale. On the other hand DPS forces the two target nucleons to be very close in transverse space. The contribution of short range correlations in the two body nuclear density may therefore give non negligible effects, considering that the value of the scale of the short range nuclear correlation is $r_c\simeq0.5$fm \cite{Dzhibuti:1975uz, Bianconi:1994vf}.

To the present purposes, a relevant property is that short range nuclear correlations are universal\cite{Alvioli:2011aa, Alvioli:2012qa}. As discussed in the Appendix B, by treating the correlation term as a perturbation, one may write

\begin{eqnarray}
\rho^{(C,2)}(r_1,r_2)&\approx&\rho^{(2)}(r_1,r_2)\bigl[1-C(r_1-r_2)\bigr]^2\cr
C(w)&=&e^{-(w^2/2r_c^2)}
\end{eqnarray}

where $\rho^{(2)}(r_1,r_2)$ is the two body nuclear density in the single particle model and for the correlation term $C(w)$ we used a Gaussian shape. For small relative distances one can approximate

\begin{eqnarray}
\rho^{(C,2)}(r_1,r_2)\Big|_{r_1\simeq r_2}&\approx&\bigl[\rho^{(1)}(r_1)\bigr]^2\bigl[1-C(w)\bigr]^2,\qquad w=r_1-r_2
\end{eqnarray}

Taking into account that the functions $f_{x_i'}(b_i)$, $f_{x_1x_2}(\beta_1-\beta_2)$ are normalized to one and that $\rho^{(2)}(r_1,r_2)$ is smooth as a function of $r_1-r_2$, the contribution to the overlap integral, in absence of short range nuclear correlations, is equal to $\int T(B)^2d^2B$.

To evaluate the terms, in the overlap integral with the nuclear correlation $C(w)$, one needs to use explicit expressions for $f_{x_i'}(b_i)$ and $f_{x_1x_2}(\beta_1-\beta_2)$. The overlap integral is most conveniently evaluated in momentum space. The term linear in $C$ is

\begin{eqnarray}
&&\int f_{x_1x_2}(\beta)f_{x_1'}(b_1)f_{x_2'}(b_2)\bigl[\rho^{(1)}(B_1,z_1)\bigr]^2\times(-2)C(B_1-B_2,\,z_1-z_2)\cr
&&\qquad\qquad\qquad\times \delta(B_1-B_2+b_1 -b_2+\beta)\,dz_1\,dz_2\,db_1\, db_2\,d\beta\,dB_1\,dB_2\cr
&&=-2\int\bigl[\rho^{(1)}(B,z)\bigr]^2dB dz\frac{1}{(2\pi)^2}\int \tilde{f}_{x_1x_2}(q)\tilde{f}_{x_1'}(q)\tilde{f}_{x_2'}(q)\tilde{C}(q)d^2q
\end{eqnarray}

where the functions with the {\it tilde} are the two-dimensional Fourier transforms in the transverse momentum space. The generalized parton distributions are known quantities. Following \cite{Frankfurt:2002ka}, we use the expression:

\begin{eqnarray}
\tilde{f}_{x'}(q)=\Bigl(1+\frac{q^2}{m_g^2}\Bigr)^{-2}
\end{eqnarray}

with $m_g^2\simeq1.1$ Gev$^2$, for $x'\approx.03$ and small $q^2$.

If the multiparton distribution in multiplicity were a Poissonian ($K=1$) and in absence of transverse correlations one would have

\begin{eqnarray}\label{eq:uncorr}
\frac{1}{\sigma_{eff}}=\int\bigl[\tilde{f}_{x_1x_2}(q)\bigr]^2\frac{d^2q}{(2\pi)^2},\quad \tilde{f}_{x_1x_2}(q)=\tilde{f}_{x_1}(q)\times\tilde{f}_{x_2}(q)=\Bigl(1+\frac{q^2}{m_g^2}\Bigr)^{-4}
\end{eqnarray}

In such a case one would obtain for the effective cross section

\begin{eqnarray}
\sigma_{eff}=\frac{28\pi}{m_g^2}=31.36\,{\rm mb}
\nonumber
\end{eqnarray}

while ATLAS measures $\sigma_{eff}=15 mb$, which implies that the uncorrelated option gives roughly an effective cross section too large by a factor two. According with Eq.(\ref{eq:seff}), $\sigma_{eff}=\pi\Lambda^2/K^2$. The uncorrelated case correspond to the values $K=1$ and $\Lambda^2=28/m_g^2$. Correlations may thus be introduced by allowing different values for $K$ and $\Lambda$, keeping however fixed their ratio in order to reproduced the measured value of the effective cross section. To have an indication on how different values of $K$ and $\Lambda$ can affect the DPS cross section in $p$\,-$A$ collisions we have considered two extreme options:

Option a) $\Lambda^2=28/m_g^2$ and $K^2=31.36/15\approx2$, which corresponds to the case where the actual value of $\sigma_{eff}$ is solely due to the multiplicity of parton pairs in the hadronic structure. In such a case the multiplicity of parton pairs would be about a factor $K\approx1.45$ times larger than expected if the distribution in multiplicity were a Poissonian, while transverse correlation between parton pairs would be completely absent, in such a way that the distribution of pairs in transverse space would be obtained by the convolution of two one-body distributions.

Option b) $K^2=1$ and $\pi\Lambda^2=15$ mb, which corresponds to assume a Poissonian for the multiparton distribution in multiplicity and to introduce a smaller typical transverse distance between partons, in compared with the uncorrelated case. The functional form of the correlated distribution of parton pairs in transverse space is unknown. One would however expect that the main features will be determined by the value of the actual scale characterizing the typical transverse distance. To proceed, we will thus consider the simplest option, where the functional form of $\tilde{f}_{x_1x_2}(q)$ is the same as in the uncorrelated case and the only modification is in the value of the scale $m_g$, which we replace with the relevant scale for the transverse separation between the parton pairs, which we denote with $h_c$. When $K^2=2$ one thus has $h_c=m_g$ while, to reproduce the observed value of $\sigma_{eff}$ when $K^2=1$, one has $h_c\approx1.52$ GeV.

The two options correspond therefore to the values:

a) $h_c^2=1.1$ GeV$^2$, $K^2=2$;

b) $h_c^2=2.3$ GeV$^2$, $K^2=1$.

By evaluating the overlap integral one obtains

\begin{eqnarray}
\frac{1}{(2\pi)^2}\int \tilde{f}_{x_1x_2}(q)\tilde{f}_{x_1'}(q)\tilde{f}_{x_2'}(q)\tilde{C}(q)d^2q={\cal C}_K\,r_c\cr
\end{eqnarray}

with

\begin{eqnarray}
{\cal C}_K&=& 0.82,\,\, {\rm in\,\, case\,\, a)}\cr
{\cal C}_K&=& 0.99,\,\, {\rm in\,\, case\,\, b)}\cr
\nonumber
\end{eqnarray}

The cross section is given by

\begin{eqnarray}\label{eq:sigma2}
\sigma_{D}^{pA}(WJJ)\big |_2&=&K\Bigl[\frac{Z}{A}\sigma_S^{pp}(W)+\frac{A-Z}{A}\sigma_S^{pn}(W)\Bigr]\sigma_S^{pp}(JJ)\\
&&\quad\times\Bigl[\int T(B)^2d^2B-2\int \rho(B,z)^2d^2Bdz\times r_c\,{\cal C}_K\Bigl]
\nonumber
\end{eqnarray}

where we made the approximation $\sigma_S^{pp}(JJ)\approx\sigma_S^{pn}(JJ)$. The ratio

\begin{eqnarray}
{\cal R}=\frac{\sigma_{D}^{pA}(WJJ)}{\sigma_{D}^{pA}(WJJ)\big |_1}
\end{eqnarray}

is thus independent on the final state phase space:

\begin{eqnarray}
{\cal R}=1+K\frac{\sigma_{eff}}{A}\Bigl[\int T(B)^2d^2B-2\int \rho(B,z)^2d^2Bdz\times r_c\,{\cal C}_K\Bigl]
\end{eqnarray}

For lead, using the Woods-Saxon nuclear density, in the two cases a) and b) one obtains

\begin{eqnarray}
{\rm a)} &&K^2=2, \pi\Lambda^2=31.36\,{\rm mb}:\quad{\cal R}=1+2.94\approx4\\
{\rm b)} &&K^2=1, \pi\Lambda^2=15\ {\rm mb}\quad\,:\quad{\cal R}=1+2.03\approx3
\nonumber
\end{eqnarray}

and the correction induced by short range nuclear correlations to the term $\int T(B)^2d^2B$ is about 8\% in case a) and about 10\% in case b). The ratio ${\cal R}$ therefore depends weakly on nuclear correlations and is rather sensitive to the different options for the values of $K$ and $\Lambda$. Notice also the strong dependence of $\sigma_{D}^{pA}(WJJ)\big |_2$ (Eq.\ref{eq:sigma2}) on $K$ and its weak dependence on $\Lambda$ (only through ${\cal C}_K$).

\section{Numerical estimates}

To have a rough estimate of the different effects of the DPS in $p$\,-$Pb$ and in $p$\,-$p$ collisions, one may compare the production rates in the same kinematical region where DPS has been measured by ATLAS in $p$\,-$p$ collisions. According with ATLAS, the fraction of events with DPS is about 7\% and one would not expect that, in interactions with a nucleus, the ratio $\sigma_{D}^{pA}(WJJ)\big |_1/\sigma_{S}^{pA}(WJJ)$ will be much different. Taking the ratio

\begin{eqnarray}
\frac{\sigma^{pA}(WJJ)}{\sigma_{S}^{pA}(WJJ)}=1+\frac{\sigma_{D}^{pA}(WJJ)\big |_1}{\sigma_{S}^{pA}(WJJ)}\times{\cal R}
\end{eqnarray}

one thus obtains that the fraction of events with DPI will grow to about 27.3\%, if there are no transverse correlations (case a), and to about 22.5\%, if the distribution in multiplicity is Poissonian (case b).

Nuclear effects and the different roles of parton correlations are of course more transparent by looking at differential distributions.

To have some indication on the differential distributions, we have evaluated the DPS differential cross section in $p$\,-$p$ and $p$\,-$Pb$ collisions, according with the different options discussed above. The elementary cross section are evaluated at the leading order in perturbation theory.
For the numerical integration we used two different sets of PDF, provided by the LHAPDF interface \cite{Whalley:2005nh}, the LO MSTW (MSTW2008lo68cl) and the CTEQ6 LO. The Leading Order matrix elements are generated by means of MadGraph 5 package \cite{Alwall:2011uj}, in the framework of the Standard Model with the CKM matrix. We choosed its C++ output and introduced a namespace characterizing every subprocess.

For the multi-dimensional integration we used VEGAS \cite{Press:2007:NRE:1403886}. More specifically we used Suave (SUbregion-Adaptive VEgas), an algorithm implemented in the CUBA library \cite{Hahn:2004fe}, which
 combines the advantages of Vegas and subregion sampling. The division into subregions allows to overcome the Vegas' problem to adapt its weight function to structures not aligned with the coordinate axes.

For a more direct comparison with available results in $p$\,-$p$, we worked out the differential distributions, both in $p$\,-$p$ and in $p$\,-$Pb$ collisions, in the same kinematical conditions of the ATLAS DPS measurements\cite{Aad:2013bjm}. Namely the beam energy is $\sqrt{s} = 7\, TeV$, $|\eta_{l^+}| < 2.47$, ${p_{tl^+}} > 20\, GeV$, $\slashed{E_t} > 25\, GeV$, ${m_t}_W > 40\, GeV$. We did not implement any fragmentation and, to reproduce the observed cross section, we slightly increased (by 10-15\%) the lower cutoff in the transverse momentum of the large $p_t$ partons\cite{Field:1989:ApQCD}. Jets are thus identified with large $p_t$ final state partons and, considering that in $p$\,-$A$ collisions the transverse spectra are not modified substantially by the presence of the nucleus\cite{ALICE:2012mj, CMS}, the effects of the nuclear modification factors are not taken into account. To simulate the process, during the integration we took trace of the final particles configurations and of the value of the integrand and plotted them with an analysis program such as ROOT \cite{root}. In the case of DPS, the transverse momentum of the boson $W^+$ was obtained following the prescriptions of \cite{Harris:1986gd, Quackenbush:2011bf, Barger:0201058766}.

\begin{figure}[h]
\centering
\includegraphics[width=70mm]{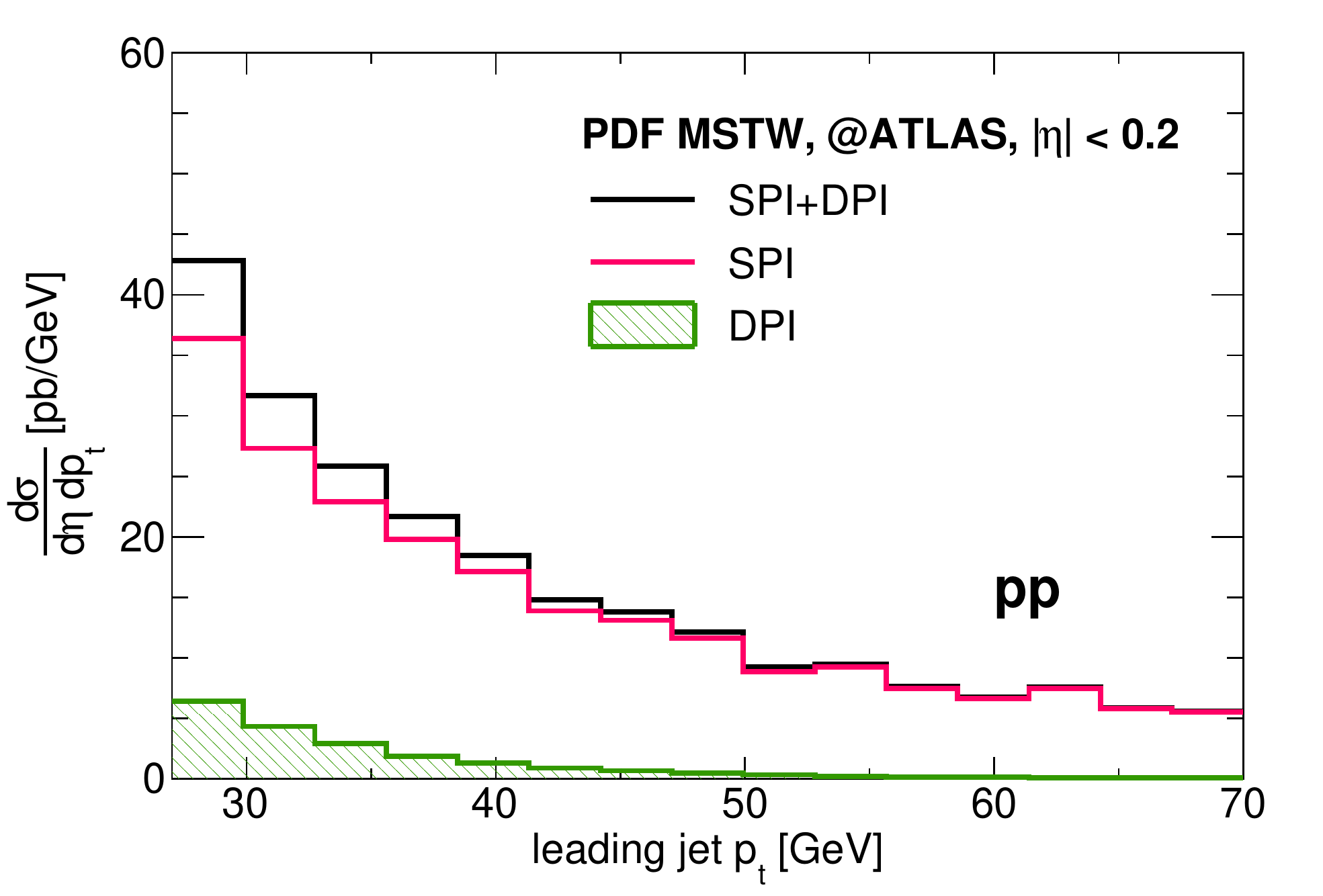}
\includegraphics[width=70mm]{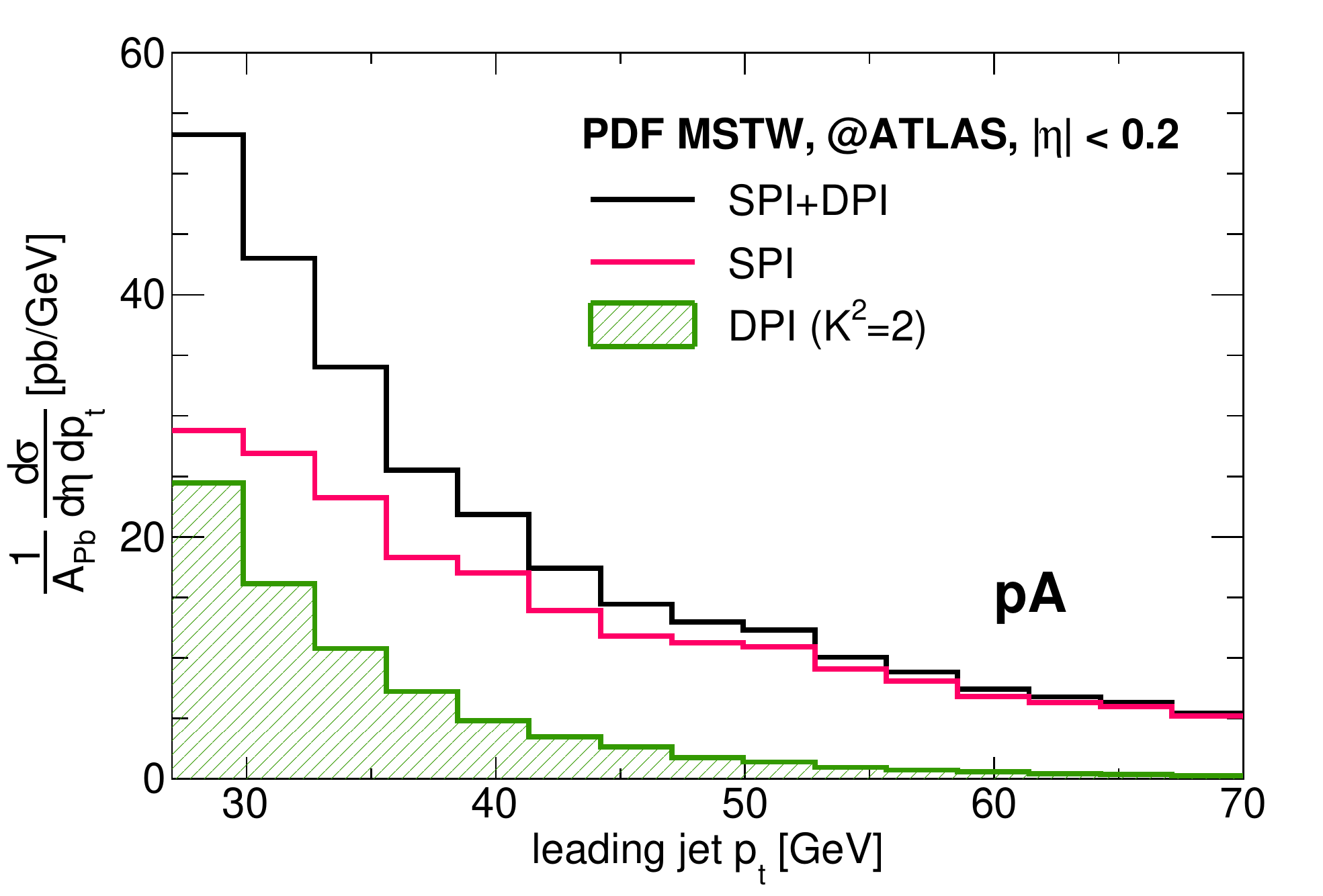}
\caption {Distribution in $p_t$ of the leading jet in $p$\,-$p$ collisions (left panel) and in $p$\,-$Pb$ collisions (right panel). The pink histograms refer to the single scattering contribution, the green ones to the DPS contribution, the black histograms are the sum of the two contributions. We used $\sigma_{eff}=15$ mb and $K^2=2$}
\label{fig:LJppA}
\end{figure}

In Fig.\ref{fig:LJppA} we plot the distribution in $p_t$ of the leading jet in $p$\,-$p$ collisions. In the DPS contribution (in green) we used $\sigma_{eff}=15$ mb (left panel). The same distribution is shown in $p$\,-$Pb$ collisions in the right panel. The pink histograms refer to the single scattering contribution, the green ones to the DPS contribution, the black histograms are the sum of the two contributions. The histograms in the figures have been computed with the MSTW parton distribution functions.

While in $p$\,-$p$ collisions DPS represent a barely noticeable contribution to the $p_t$ spectrum of the leading jet produced in the process, DPS have a much stronger effect in the $p_t$ spectrum of the leading jet in $p$\,-$Pb$ collisions, where the shape of the distribution is very different for $p_t$ smaller than 50 GeV.

\begin{figure}[h]
\centering
\includegraphics[width=70mm]{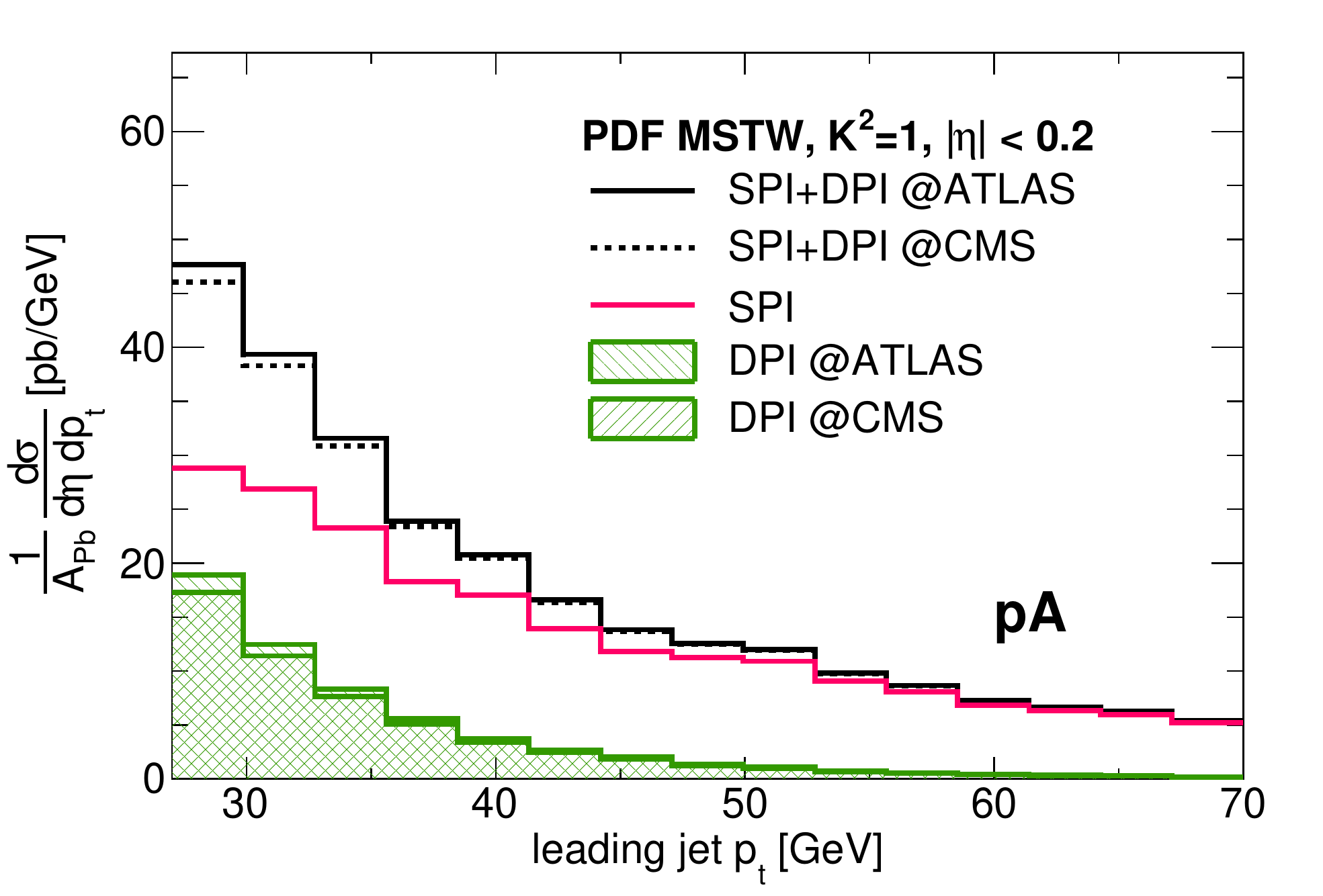}
\includegraphics[width=70mm]{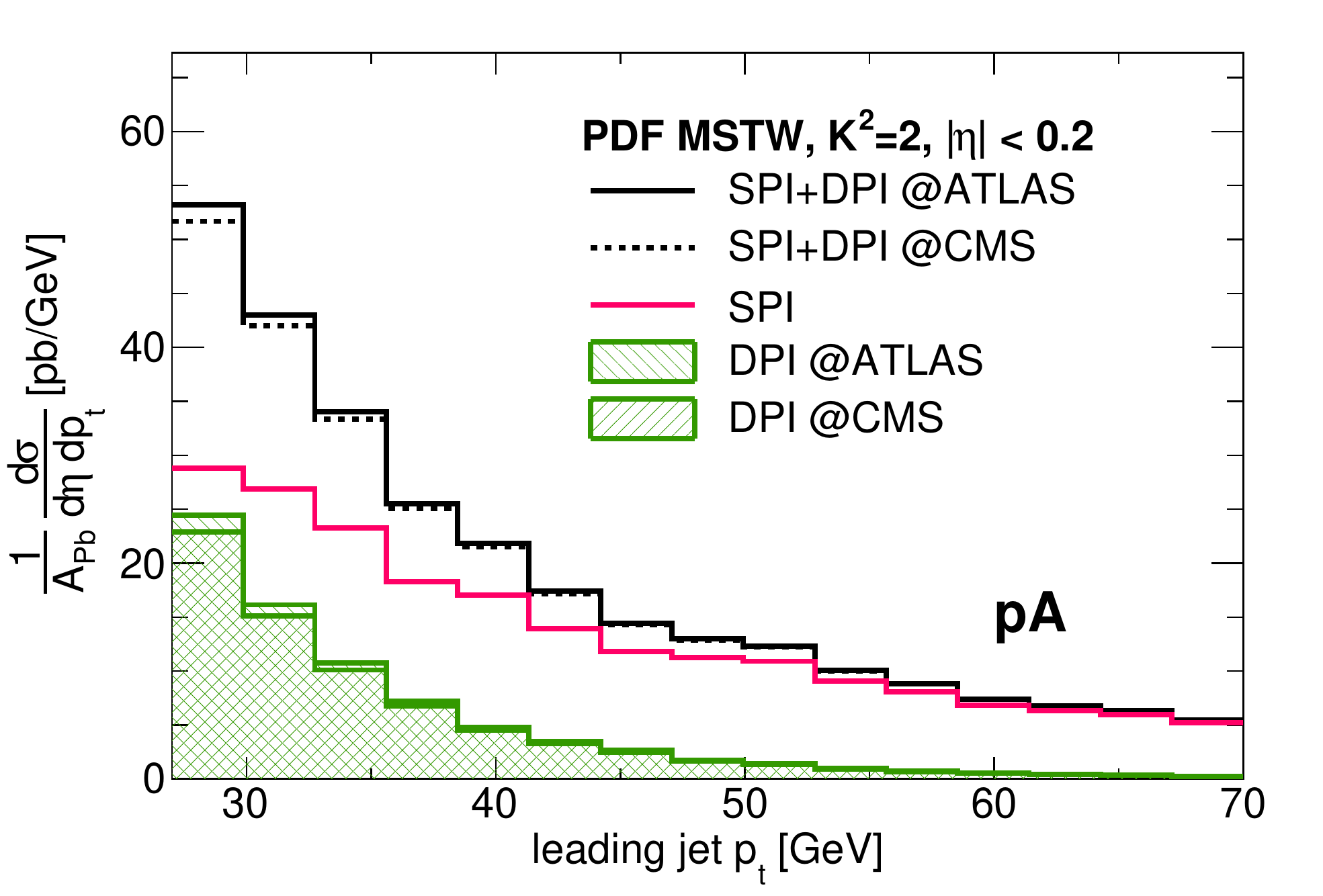}
\caption {Distribution in $p_t$ of the leading jet in $p$\,-$Pb$ collisions in the cases $\sigma_{eff}=15$ mb (ATLAS) and $\sigma_{eff}=20.7$ mb (CMS\cite{CMS:awa}) for $K^2=1$ (left panel) and  and $K^2=2$ (right panel)}
\label{fig:LJCMSATLAS}
\end{figure}

The dependence of the transverse spectrum of the leading jet, as a function of the value of the $\sigma_{eff}$ and of $K$, is shown in Fig.\ref{fig:LJCMSATLAS}. By looking at the green histograms one may see that, after subtracting the single scattering contribution, which can be be considered as a known quantity, once DPS have been measured in $p$\,-$p$ collisions in the same kinematical conditions, the shape in $p_t$ shows an appreciably dependence on the value of $K$.

\begin{figure}[h]
\centering
\includegraphics[width=70mm]{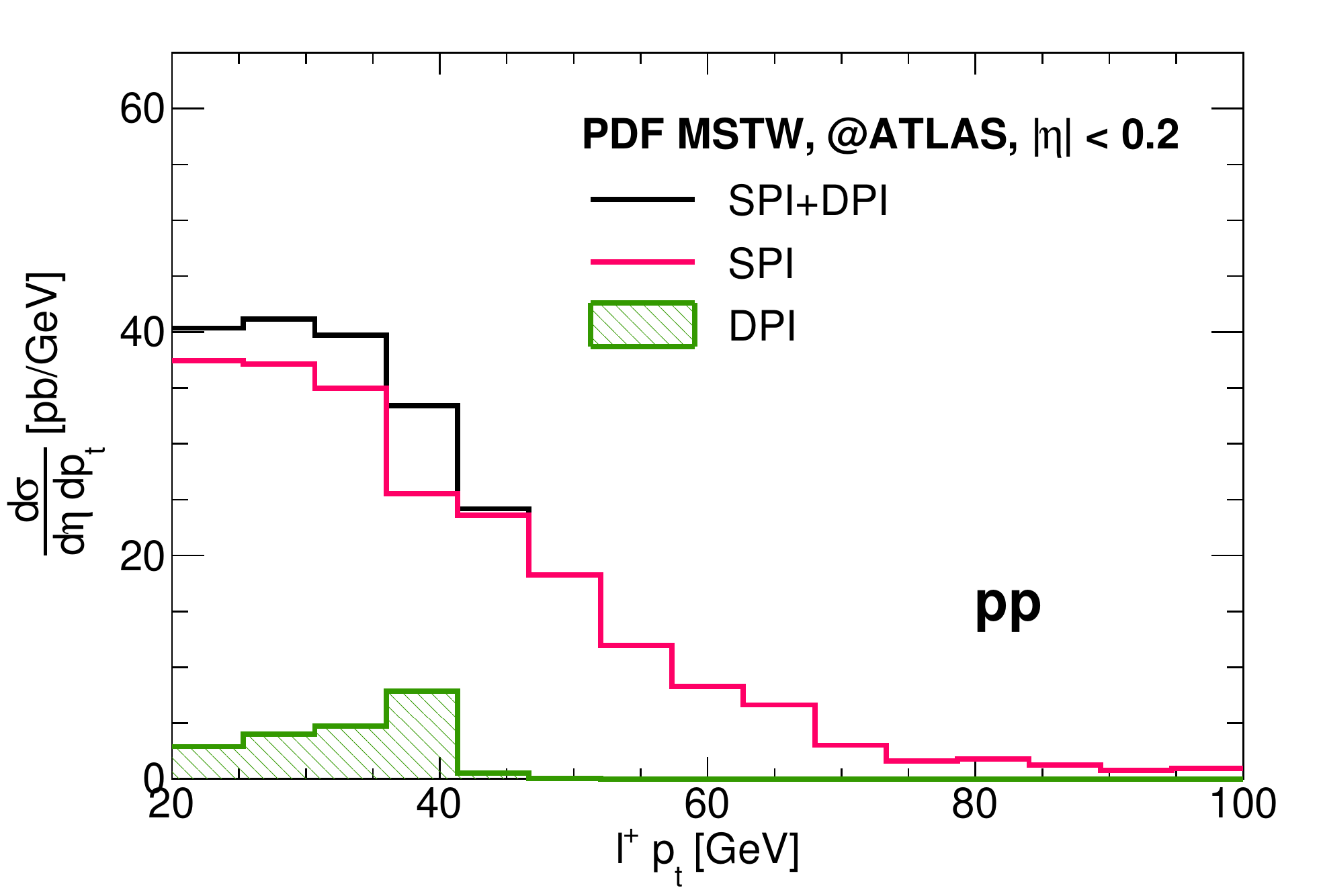}
\includegraphics[width=70mm]{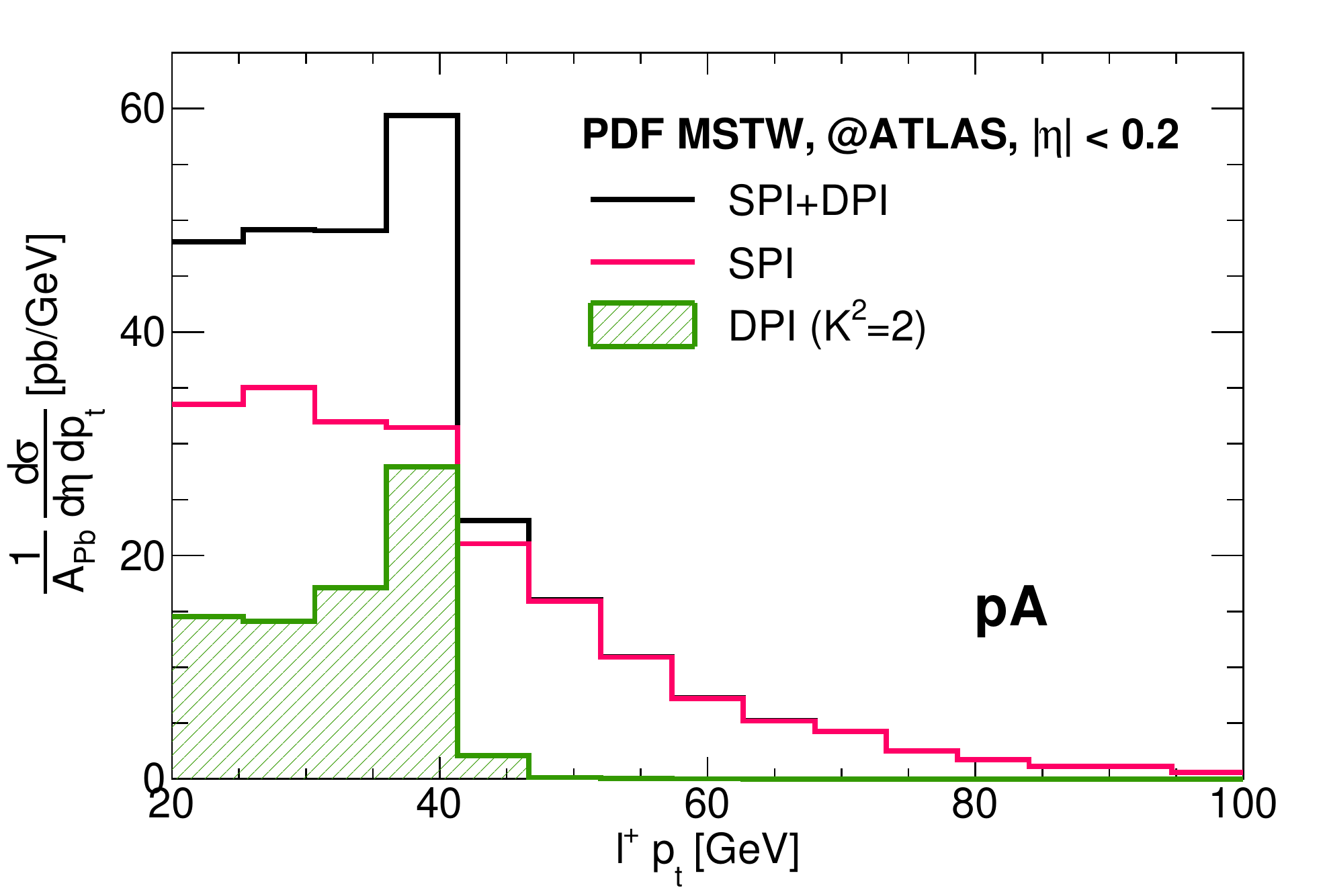}
\caption {Distribution in $p_t$ of the charged lepton from the $W^+$ decay, in $p$\,-$p$ (left panel) and in $p$\,-$Pb$ collisions in the case of $\sigma_{eff}=15$ mb and $K^2=2$ (right panel)}
\label{fig:LeptonppA}
\end{figure}

A more suitable observable, to learn on the distribution in multiplicity of the multi-parton distribution, is probably the $p_t$ spectrum of the charged lepton, produce by the decay of the $W^+$. In a single scattering collision $W$ bosons recoil against the produced  jets and are typically characterized by a transverse momentum of the order of the lower cutoff in $p_t$ of the observed accompanying jets. In the case of a DPS, the jets and the $W$ are produced in different partonic interactions. The transverse momentum of the $W$ is therefore typically rather small and the spectrum of the decay lepton is thus rather different in single and in double parton scattering. In the former case, when the lower cutoff for the produced jets is 20 GeV, the transverse momentum of the produced lepton can easily exceed 60-70 GeV. In the latter case the lepton is produced by a $W$ boson with a rather small transverse momentum and its transverse spectrum is thus limited to values close to $1/2$ of the $W$ mass. In Fig.\ref{fig:LeptonppA} we plot the distribution in $p_t$ of the charged lepton from the $W^+$ decay. The left panel refers to the case of $p$\,-$p$ collisions; the right panel to the case of $p$\,-$Pb$ collisions. The enhancement of the spectrum at $p_t<40$ GeV, due to the contribution of DPS, is not a big effect in $p$\,-$p$ collisions. It is on the contrary a rather strong effect in  $p$\,-$Pb$ collisions, where the difference with respect to the contribution to the spectrum due to single parton scattering (pink histograms in Fig.\ref{fig:LeptonppA}) is quite noticeable.

\begin{figure}[h]
\centering
\includegraphics[width=70mm]{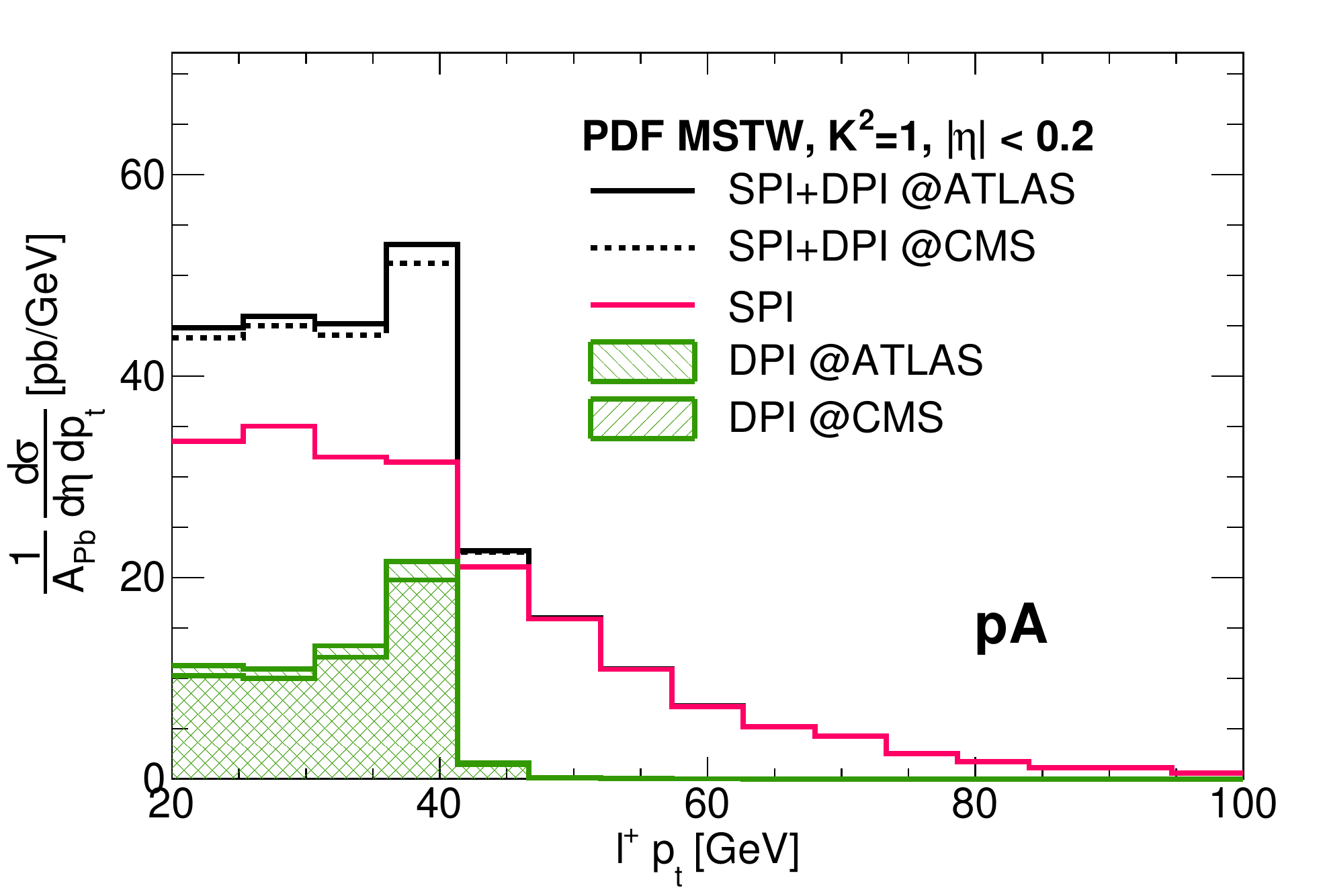}
\includegraphics[width=70mm]{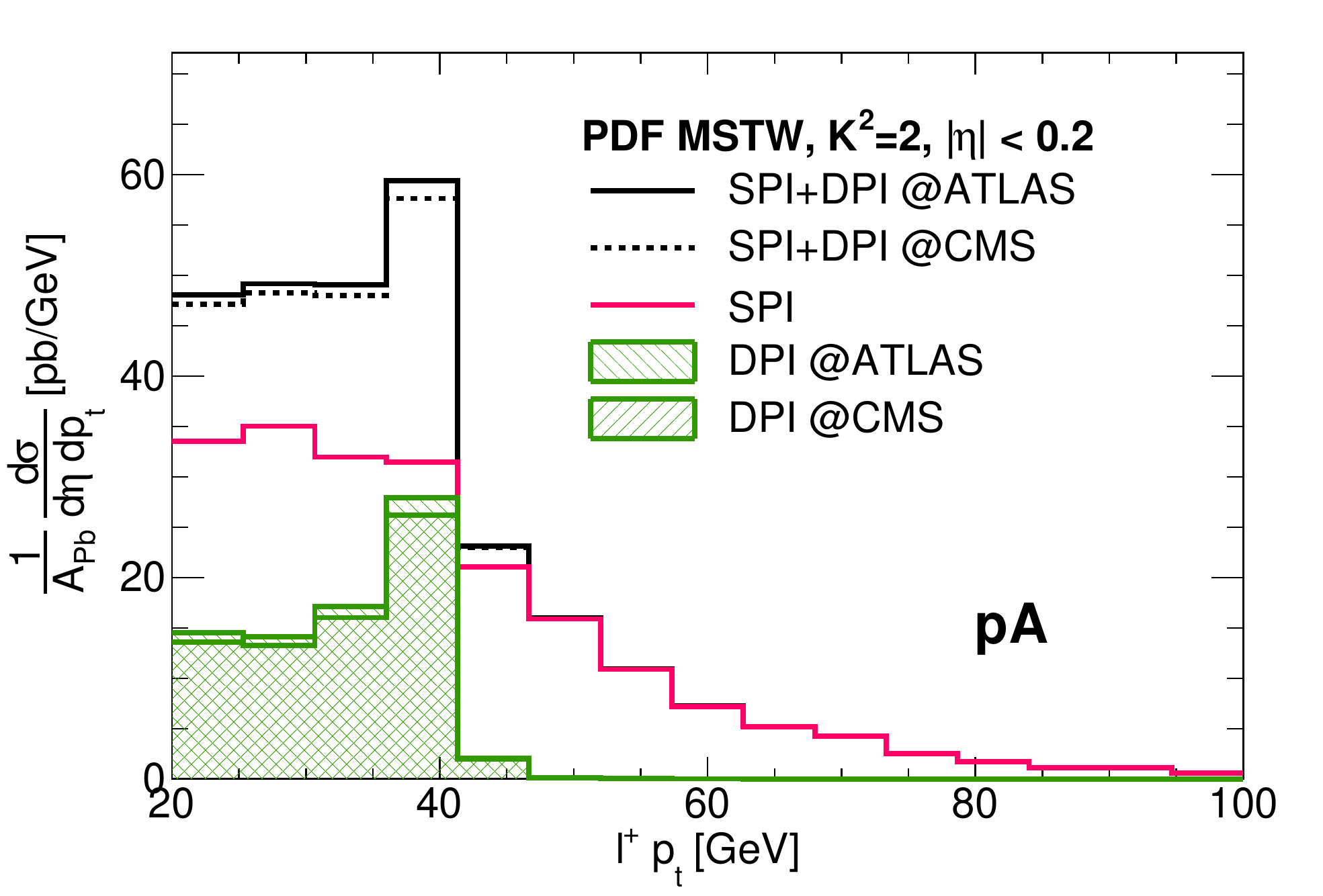}
\caption {Rapidity distribution of the charged lepton from the $W^+$ decay, in $p$\,-$Pb$ collisions in the case of $\sigma_{eff}=15$ mb (ATLAS) and $\sigma_{eff}=20.7$ mb (CMS) for $K^2=1$ (left panel)  and $K^2=2$ (right panel)}
\label{fig:LeptonCMSATLAS}
\end{figure}

In $p$\,-$Pb$ collisions, the sensitivity of the spectrum to $K$ is shown in Fig.\ref{fig:LeptonCMSATLAS}. In the left panel we show the spectrum in the case $K^2=1$ for $\sigma_{eff}=15$ mb and $\sigma_{eff}=20.7$ mb. In the right panel we show the case $K^2=2$. The enhancement of the spectrum due to the DPS contribution at $p_t<40$ GeV is rather substantial and the amount of the increase is significantly different as a function of $K$.

By selecting events with a charged lepton in the $40$ GeV $p_t$ region one will thus obtain a sample where the contribution of DPS is of about 50\% of the total and one will thus be able to obtain a rather direct information on the second moment of the multi-parton distribution in multiplicity in the proton structure.

A final observation is that, due to the different production mechanism as compared to the case of $p$\,-$p$ collisions, nuclear spectra do not depend much on the value of $\sigma_{eff}$ measured in $p$\,-$p$ collisions. A change from 20 to 15 mb implies an increase of the DPS cross section of more than 30\% in $p$\,-$p$ collisions and of only 5-6\% in $p$\,-$Pb$ collisions, as apparent in the figures above by comparing the dotted and continuous histograms. In the latter case the production rate is in fact proportional, to a large extent, to the multiplicity of pairs of partons in the projectile, while the typical transverse distance between the interacting parton pairs does not play a relevant role. Which is precisely the reason why DPS in $p$\,-$Pb$ collisions have the potential to provide a lot of information on parton correlations.

\section{Concluding Summary}

Double Parton Scattering processes are directly related to unknown non perturbative properties of the hadron structure, which in $p$\,-$p$ collisions converge in the value of a single quantity, the effective cross section. The interaction mechanism is more complex in $p$\,-$A$ collisions, where one may have either one or two different target nucleons interacting with large momentum transfer. In the case of two different active target nucleons, in addition to the diagonal contribution, which has a direct probabilistic interpretation, one may need to take into account also the contribution of an interference term. When the two active partons in the initial state are identical, the nucleus can in fact generate the initial partonic configuration in two different ways. The description of the interaction is simpler when the interference term is absent, which is the case of reaction channels where the pair of initial state active partons are a quark and a gluon, like in $WJJ$, $Wb{\bar b}$ and $Wc{\bar c}$ production at relatively low fractional momenta. The increased complexity of the interaction in $p$\,-$A$ collisions can thus provide an additional handle to obtain information on the non-perturbative hadron structure not accessible by other means.

To gain some insight into the actual possibilities of learning about parton correlations by studying DPS in $p$\,-$p$ and $p$\,-$A$ collisions, we have considered a particularly simple case, still not inconsistent with present experimental evidence, where DPS is described by the dominant term at small $x$ while, in $p$\,-$p$ collisions, the effective cross section can be approximated with a universal constant. The effective cross section is thus fully determined by the typical transverse distance between the interacting partons $\Lambda$ (Eq.\ref{eq:Lambda}) and by the multiplicity of parton pairs, which here is characterized by the value of $K$ (Eq.\ref{eq:kappa}). To keep the interaction with the nucleus as simple as possible, we have looked at a reaction channel where the interference term is absent. Specifically we have studied $WJJ$ production, which is of particular interest since DPS in $WJJ$ production is presently studied experimentally in $p$\,-$p$ collisions both by ATLAS and by CMS.

In this simplified scheme, the DPS cross section in $p$\,-$p$ collisions depends only on the ratio between $K$ and $\Lambda$. In $p$\,-$A$ collisions the contribution to the DPS cross section with two active target nucleons depends on the contrary (almost) only on $K$. To have an indication on the possibility of determining $\Lambda$ and $K$, by measuring DPS in $p$\,-$p$ and $p$\,-$Pb$ collisions and to allow a direct comparison of the two cases, we have evaluated the $W^+JJ$ production cross sections, in the kinematical conditions of the ATLAS experiment.

In $p$\,-$A$ collisions, the contribution to the DPS cross section, due to the processes where two different target nucleons interact with large momentum transfer, is proportional to the factor $K$ and grows with $A^{4/3}$. Depending on the value of $K$, in $p$\,-$Pb$ collisions this contribution may be twice or three times as big as the contribution to DPS, where only a single target nucleon interacts with large momentum transfer, while short range nuclear correlations can produce at most a reduction of 10\%. The effect of varying the value of the typical distance in transverse space, between the pairs of interacting partons, has in this case only a minor effect, which we estimate to be of the order of 5-6\% of the cross section. Considering also the contribution due to single hard collisions we expect that, while in $p$\,-$p$ the observed fraction of events with DPS was about 7\%, with the same cuts used by ATLAS, in $p$\,-$Pb$ the fraction of events with DPS will range between 22.5\% (in the absence of longitudinal correlations) and 27.3\% (in the absence of transverse correlations). CMS reports a smaller fraction of events with DPS, about 5\%. In such a case and with the same cuts, in $p$\,-$Pb$ collisions the fraction of events with DPS will range between 21.0\% and 26.0\%.

In summary, a main feature of DPS in $p$\,-$A$ collisions is that, for large atomic mass numbers, the most important contribution to the DPS cross section is due to the interactions with two active target nucleons, rather than to the interactions with a single target nucleon\cite{Strikman:2001gz}\cite{Cattaruzza:2004qb}. In other terms DPS in $p$\,-$Pb$ collisions is characterized by a very strong anti-shadowing, which may represent a 200-300\% correction to the DPS cross section on a single nucleon. The amount of this anti-shadowing term is proportional to the flux of incoming pairs of partons and, by measuring the amount of anti-shadowing, one has thus a direct indications on the number of pairs of incoming partons in the projectile. Different properties of the incoming pair of partons, which may have important effects in DPS in $p$\,-$p$ collisions, are likely to be much less important in $p$\,-$A$. In the actual case we have discussed the effect of the distribution of partons in transverse space, which in $p$\,-$p$ may be even more important than the multiplicity of parton pairs, to determine the observed value of the DPS cross section. In $p$\,-$Pb$, we estimate that different values of the typical separation between partons in transverse space can, on the contrary, affect the DPS cross section only by about 5-6\%.

We think that, although there are still several open problems to understand DPS in $p$\,-$p$ collisions, the study of DPS in $p$\,-$A$ collisions has therefore a great potential for a deeper inside in the problem. DPS in $p$\,-$A$ collisions allow in fact to single out an important feature in the process, actually the value of incoming flux of parton pairs, which is directly proportional to the amount of anti-shadowing observed in the DPS cross section. In the present paper we have worked out the amount of anti-shadowing to be expected in the simplest conceivable scheme and the corresponding value of the incoming flux of parton pairs. A comparison with an experimental study of DPS in $p$\,-$Pb$ at the LHC would thus be very instructive, providing for the first time a direct indication on a property of the correlated parton structure of the hadron, not achievable to our knowledge by other means and allowing, at the same time, to make a quantitative test or even to disprove the simplest conceivable description of DPS.

\vskip.25in

\appendix

\section{\bf Diagonal and interference terms}

\vskip.25in

 The construction of the amplitudes and of the corresponding cross sections,
 Eq.s (\ref{eq:correlation}, \ref{eq:offdiag})
in the main text, is performed by following strictly the procedure used in \cite{Treleani:2012zi},
 in particular in the discussion of the Tritium case. We will not reproduce all details of the procedure. Rather we will try to point out some of the main differences between the two cases.

\begin{figure}[h]
\centering
\includegraphics[width=140mm]{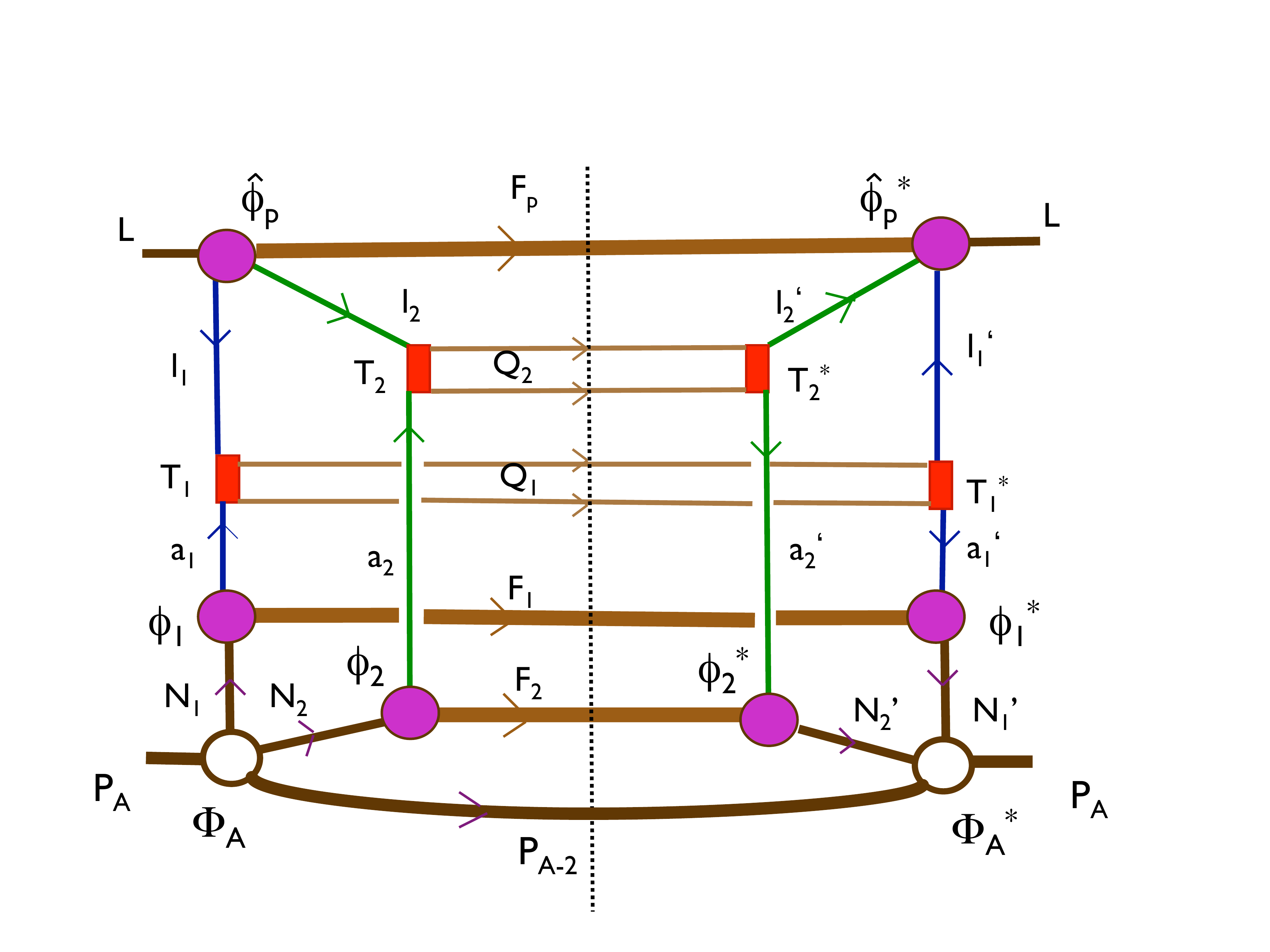}
\caption {Discontinuity of the forward amplitude in Eq.(\ref{eq:Appdigonal})}
\label{fig:Appdigonal}
\end{figure}

 \par
  The diagonal term, Fig.\ref{fig:Appdigonal}, is given by:

\begin{eqnarray}
\label{eq:Appdigonal}
{\rm Disc}&{\cal A}_d &=\frac{1}{(2\pi)^{21}}\int\frac{{\hat\phi}_p}{{l_1}^2 {l_2}^2}\ \frac{{\hat\phi_p}^*}{{l'_1}^2 {l'_2}^2}\ \frac{\phi_1}{a_1^2}\;
\frac{\phi_1^*}{{a'_1}^2}\;\frac{\phi_2}{{a_2}^2}\;\frac{\phi_2^*}{{a'_2}^2}\nonumber\\
&\times&T_1(l_1, a_1\to q_1,q'_1)\; T_1^*(l'_1,a'_1\to q_1,q'_1)\;T_2(l_2, a_2\to q_2,q'_2)\;T_2^*(l'_2, a'_2\to q_2,q'_2)\nonumber\\
&\times&\frac{\Phi_A(N_1;N_2|N_k)}{[N_1^2-m^2][N_2^2-m^2]}\frac{\Phi_A^*(N'_1;N'_2|N_k)}{[{N'_1}^2-m^2][{N'_2}^2-m^2]}\nonumber\\
&\times&\delta(L-l_1-l_2-F_p)\;\delta(L-l'_1-l'_2-F_p)\nonumber\\
&\times&\delta(N_1-a_1-F_1)\;\delta(N'_1-a'_2-F_1)\;\delta(N_2-a_2-F_2)\;\delta(N'_2-a'_1-F_2)\nonumber\\
&\times&\delta(l_1+a_1-Q_1)\:\delta(l'_1+a'_1-Q_1)\;\delta(l_2+a_2-Q_2)\;\delta(l'_2+a'_2-Q_2)
\nonumber\\
& \times&\prod_{i,j} \;d^4a_i d^4a'_i d^4l_i d^4l'_i d^4 F_j\delta ({F_j}^2-{M_j}^2) \delta(P_A-N_1-N_2-P_{A-2})\delta(P_A-N'_1-N'_2-P_{A-2})\nonumber\\
&\times&\delta\Bigl(\sum_{k=3}^AN_k-P_{A-2}\Bigr)\prod_{k=3}^Ad^4N_kd^4 P_{A-2} d^4N_1 d^4N_2 d^4N'_1 d^4N'_2d^4 Q_i
d(\Omega_i/8)d{M_j}^2
\label{two}
\end{eqnarray}

The interference term, Fig.\ref{fig:Appinterf}, is given by:

\begin{figure}[h]
\centering
\includegraphics[width=140mm]{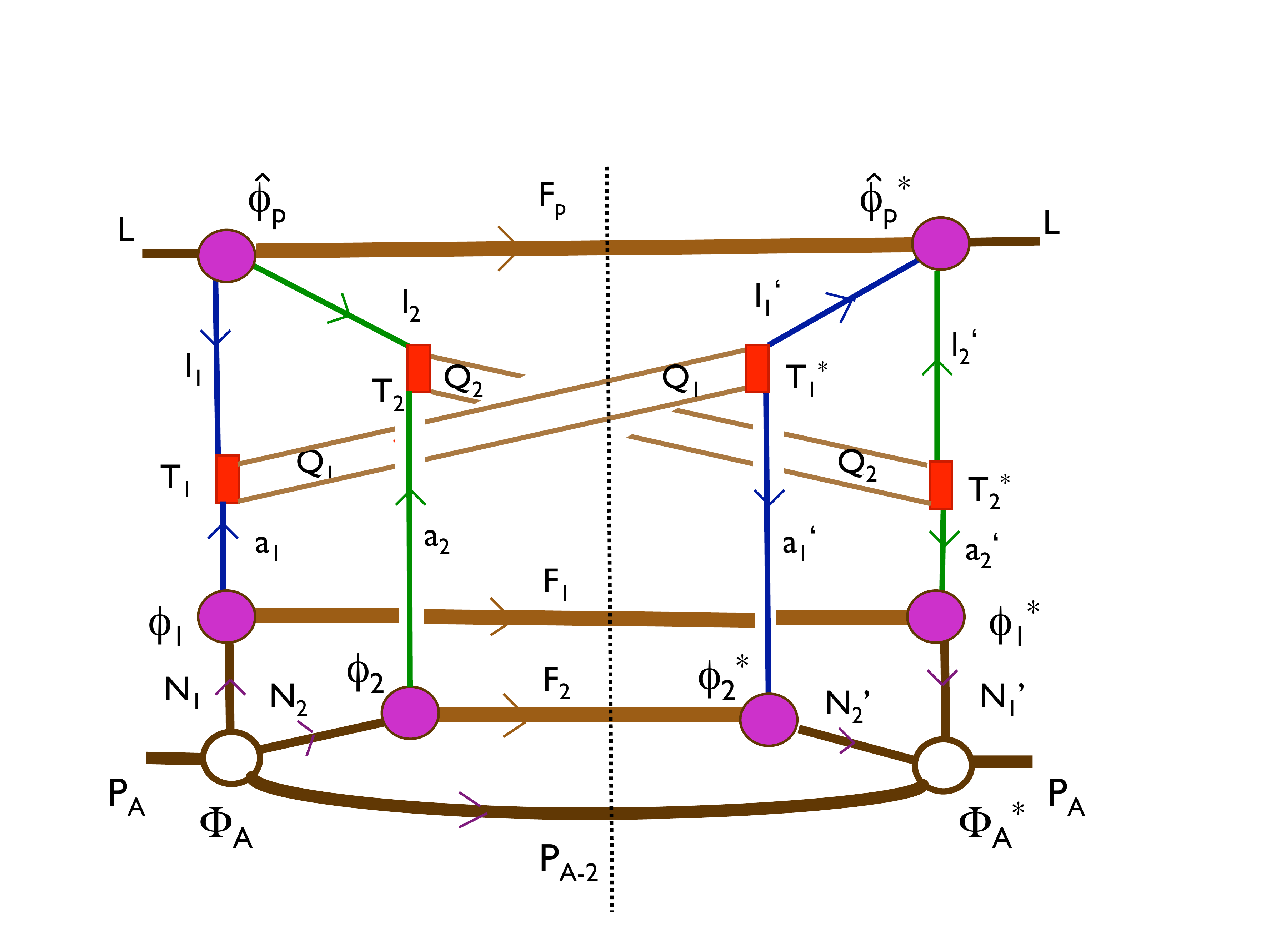}
\caption {Discontinuity of the forward amplitude in Eq.(\ref{eq:Appinterf})}
\label{fig:Appinterf}
\end{figure}

\begin{eqnarray}
\label{eq:Appinterf}
{\rm Disc}&{\cal A}_i& =\frac{1}{(2\pi)^{21}}\int\frac{{\hat\phi}_p}{{l_1}^2 {l_2}^2}\ \frac{{\hat\phi_p}^*}{{l'_1}^2 {l'_2}^2}\ \frac{\phi_1}{a_1^2}\;
\frac{\phi_2^*}{{a'_1}^2}\;\frac{\phi_2}{{a_2}^2}\;\frac{\phi_1^*}{{a'_2}^2}\nonumber\\
&\times&T_1(l_1, a_1\to q_1,q'_1)\; T_1^*(l'_1,a'_1\to q_1,q'_1)\;T_2(l_2, a_2\to q_2,q'_2)\;T_2^*(l'_2, a'_2\to q_2,q'_2)\nonumber\\
&\times&\frac{\Phi_A(N_1;N_2|N_k)}{[N_1^2-m^2][N_2^2-m^2]}\frac{\Phi_A^*(N'_1;N'_2|N_k)}{[{N'_1}^2-m^2][{N'_2}^2-m^2]}\nonumber\\
&\times&\delta(L-l_1-l_2-F_p)\;\delta(L-l'_1-l'_2-F_p)\nonumber\\
&\times&\delta(N_1-a_1-F_1)\;\delta(N'_1-a'_2-F_1)\;\delta(N_2-a_2-F_2)\;\delta(N'_2-a'_1-F_2)\nonumber\\
&\times&\delta(l_1+a_1-Q_1)\:\delta(l'_1+a'_1-Q_1)\;\delta(l_2+a_2-Q_2)\;\delta(l'_2+a'_2-Q_2)\nonumber\\
& \times&\prod_{i,j} \;d^4a_i d^4a'_i d^4l_i d^4l'_i d^4 F_j\delta ({F_j}^2-{M_j}^2) \delta(P_A-N_1-N_2-P_{A-2})\delta(P_A-N'_1-N'_2-P_{A-2})\nonumber\\ &\times&\delta\Bigl(\sum_{k=3}^AN_k-P_{A-2}\Bigr)\prod_{k=3}^Ad^4N_k d^4 P_{A-2} d^4N_1 d^4N_2 d^4N'_1 d^4N'_2d^4 Q_i
d(\Omega_i/8)d{M_j}^2
\end{eqnarray}

Here $P_A$ is the four-momentum of the incoming nucleus and $L$ the four-momentum of the incoming proton; $\phi$ and $\hat \phi$ are the effective vertices for emission of one or two partons by a nucleon; the integration variables $Q$ and $\Omega$ come in through the transformation
$d^3q/2q_o d^3q'/2q'_o=d^4Qd\Omega/8$, where $q$ and $q' $ are the momenta of the massless particles produced in the hard scattering $T_i$, so $Q^2>0,\;Q_o>0$ and $\Omega$ gives the scattering angles in center-of-momentum frame of the pair.
The sum is over every possible final state compatible with the conservation laws, identified by the four-vectors $N_k$, $(k=3,\dots A)$.\\
  We recall\cite{Treleani:2012zi}
the amplitude for finding one or two partons in the projectile when the remnant of the parent nucleon has mass $M_j$, where the possible values of the index are $j=1,2,p$ (cf. figures \ref{fig:Appdigonal} and \ref{fig:Appinterf}); setting $\lambda_-=\frac{1}{2}(l_1-l_2)_-$ we get:

\begin{eqnarray}
 \psi_{M_{j=1,2}}&=&\frac {\phi_j}{a_j^2}=\frac{\phi_j}{\bar x_j[m_j^2-M^2_{j\bot}/(1-\bar x_j)]-a^2_{j\bot}}\;,\cr
 \hat\psi_{M_p}&=&\frac{1}{\sqrt 2}\int\frac{\hat\phi_p}{l_1^2 l_2^2}\frac{d\lambda_-}{2\pi i}  =\frac {1}{\sqrt{2} L_-}\frac{\hat\phi_p}{{l_1}_{\bot}^2 x_2+{l_2}_{\bot}^2 x_1-
 x_1x_2[m^2-M_{p\bot}^2/(1-x_1-x_2)]}\;.
  \end{eqnarray}

 Here the light cone components that grow with the total energy in the c.m. of the interacting nucleon pair are the $+$ components in the projectile proton and the $-$ components in the target nucleon.
 The dependence on the transverse mass of the remnant: $M^2_{j\bot}\equiv M_j^2+F^2_{j\bot}$ comes in through the conservation of the $plus-$components, when $j=1$ or 2, and of the $minus-$components, when $j=p$.
 In the same way one defines the one-parton and the two-parton amplitudes in the bound nucleon
 (in this case, since $N$ has also transverse components, the initial state mass $m^2$ has been replaced by $m_j^2\equiv m^2+N_{j\bot}^2$). Then the Fourier transformation on the transverse variables is performed.\\
  The same procedure is applied to the nucleus side: one defines
$\nu_+=\frac{1}{2}(N_1-N_2)_+$ and the covariant amplitude for finding two nucleons in the nucleus has the formal expression:
\begin{eqnarray}
\Psi_A({N_1}_-,{N_2}_-)=\frac{1}{\sqrt 2}\int \frac {d \nu_+}{2\pi i}\frac {\Phi_A}{[N_1^2-m^2]\cdot [N_2^2-m^2]}
\end{eqnarray}
 \par
This amplitude depends also on the configuration of the residual $(A-2)$-nucleons.
\begin{eqnarray}
\Psi_A({N_1}_-,{N_2}_-)=\frac{1}{\sqrt 2{P_A}_-}\frac{\Phi_A}{Z_1Z_2\Bigl\{(M_A/A)^2-\mu_{\bot}^2/[A(A-Z_1-Z_2)]\Bigr\}-(Z_1 m_2^2+Z_2m_1^2)/A}\;
\end{eqnarray}
 $\mu_{\bot}$ is the overall transverse mass of the remnant nuclear spectators, $M_A$ is the mass of the incoming nucleus and ${N_i}_-=Z_i{P_A}_-/A$. In the dominant configurations $Z_i$ is close to 1. 
 \par
We try now to point out the differences between diagonal and the interference term.\\
In the diagonal term, the conservation of the large momentum components implies the following relations, for the initial state partons on the two sides of the diagram:
${l_i}_+={l'_i}_+,\;{a_i}_-={a'_i}_-,\;{N_i}_-={N'_i}_-$, while the corresponding transverse variables become diagonal through Fourier transformation. The whole expression of the cross section can thus be expressed in terms of densities $i.e.$ square of partonic wave functions and of the wave function of the bound nucleons.\\
In the interference term, the conservation of the large momentum components implies different relations, for the initial state partons on the two sides of the diagram. Actually: ${l_i}_+={l'_i}_+,\;a_{1-}=a'_{2-},\;a_{2-}=a'_{1-}$. For the nuclear variables one obtains:
$(N_1-N'_1)_-=(N'_2-N_2)_-=(a_1-a_2)_-$. Concerning the transverse components, the variables $\beta_{i\bot}$ (conjugated to  $l_{i\bot}$) become diagonal through Fourier transformation while, differently from the case of the diagonal contribution to the cross section, the variables $b_{i\bot}$ (conjugated to $a_{i\bot}$) are not diagonalized by the Fourier transformation.
When looking to the nuclear part, the overall conservation of the fractional $minus-$component implies: $Z_1+Z_2=Z'_1+Z'_2=A-Z_{A-2}$, where $Z_{A-2}$ is the $minus-$fractional momentum of the incoming nucleus. Notice that the relation for the nuclear fractional momenta can be written also as $Z_1-Z_1'=Z_2'-Z_2=\bar x_2-\bar x_1$, which shows that such differences can be actually measured.\\
While ${\cal A}_d$ can be expressed through the diagonal terms of the two-body nuclear density matrix, ${\cal A}_i$ requires the off diagonal two body density matrix. In this latter case the non perturbative partonic input of the projectile proton is given again by the partonic densities
$\Gamma(x_1,x_2;\beta_1,\beta_2)$, the non perturbative input of the target nucleons depends however explicitly on the non diagonal one-body parton densities, which, in the final expression of the cross section, are gathered into the function $W$, whose expression is shown here below:\\

\begin{eqnarray}
 && W(Z_1,Z_2;\bar x_1;\bar x_2;b_1,b_2)=\frac {1}{4(2\pi)^6} \int dM_1^2dM_2^2
 \frac {\bar x_1\bar x_2}{(Z_1-\bar x_1)(Z_2-\bar x_2)}        \cr
  && \times \psi_{M_1}(\bar x_1/Z_1,b_1) \psi_{M_2}(\bar x_2/Z_2,b_2) \psi^*_{M_2}(\bar x_2/Z'_1,b_1-B_1) \psi^*_{M_1}(\bar x_1/Z'_2,b_2-B_2)
  \end{eqnarray}

\section{\bf Two-body nuclear density}

\vskip.25in

In this appendix we describe the approach used to derive the two-body nuclear density in the main text.
Nuclear states are normalized to one: $\int\psi_m(u)^*\psi_n(u) du=\delta_{mn}$ and with $u$ we mean all nucleon's degrees of freedom, spin, isospin and space coordinates. The one-body nuclear density
$\rho^{(1)}(u)=\sum_n|\psi_n(u_1)|^2$ is normalized to the atonic mass number
$\int \rho^{(1)}(u) du=A.$
The antisymmetric two-body wave function, neglecting interactions between the two nucleons, is:

$$\frac{1}{\sqrt 2}\bigl[\psi_m(u_1)\psi_n(u_2)-\psi_m(u_2)\psi_n(u_1)\bigr]$$

\par
Correspondingly the two-body density for the states
$m,n$ is:

$$g_{mn}(u_1,u_2)=
\frac{1}{2}\big[|\psi_m(u_1)\psi_n(u_2)|^2+|\psi_m(u_2)\psi_n(u_1)]|^2\big]-
\Re\big[\psi_m(u_1)\psi_n(u_2)\psi_m(u_2)^*\psi_n(u_1)^*\big]$$

\par
By integrating on $u_2$ one obtains:

$$\int g_{mn}(u_1,u_2)du_2=
\frac{1}{2}\bigl[|\psi_m(u_1)|^2+|\psi_n(u_1)|^2\bigr]-\delta_{mn}|\psi_m(u_1)|^2$$

and by further integrating on $u_1$ the result is $1-\delta_{mn}$, which implies that summing the two-body density over $m,n=1\dots A$ one obtains $A(A-1)$.
The two body density is thus given by

$$\rho^{(2)}(u_1,u_2)=\sum_n|\psi_n(u_1)|^2\sum_n|\psi_n(u_2)|^2-
|\Delta(u_1,u_2)|^2$$

where $\Delta(u_1,u_2)=\sum_n\psi_n(u_1)^*\psi_n(u_2)$. Notice that, once $\sum_n$ is done over a complete set of states in Hilbert space, one obtains $\Delta \to \delta$.
As a direct consequence of its definition one has  $\int|\Delta(u_1,u_2)|^2du_2=\rho^{(1)}(u)$ and thus
$ \int\rho^{(2)}(u_1,u_2)du_2=(A-1)\rho^{(1)}(u_1)$.

We are interested in short range nuclear correlations, which exhibit a universal behavior\cite{Alvioli:2011aa}. We introduce therefore an $m$ and $n$ independent correlation term in the two-body wave function:

$$\frac{1}{\sqrt{2}}\bigl[\psi_m(u_1)\psi_n(u_2)-\psi_m(u_2)\psi_n(u_1)\bigr]\times
\bigl[1-C(u_1,u_2)\bigr]$$

The corresponding two body density of the states $m,n$ is

$$f_{mn}(u_1,u_2)=g_{mn}(u_1,u_2)\times
\bigl[1-C(u_1,u_2)\bigr]^2$$

\par
By summing over states one obtains the correlated two-body density

$$\rho^{(C,2)}(u_1,u_2)=\rho^{(2)}(u_1,u_2)\bigl[1-C(u_1,u_2)\bigr]^2$$

At first order in $C$  one has

$$ \int\rho^{(C,2)}(u_1,u_2)du_2=(A-1)\rho^{(1)}(u_1)-
2\int\rho^{(2)}(u_1,u_2)C(u_1,u_2)du_2$$

One should now keep into account that $u$ includes also the spin variables
$\int du\equiv \int d^3r \sum_s$. When the two nucleons are in a spin triplet state, their space wave function is antisymmetric and therefore it vanishes for $r_1\to r_2$ irrespectively of the presence of the correlation term. When the two nucleons are in a spin singlet state, their space wave function is symmetric and the effect of the short range correlation term in this case is particularly important. In the spin singlet case and without interaction, the space wave function is:

$$\Psi=\frac{1}{\sqrt{2}}\bigl[\psi_m(r_1)\psi_n(r_2)+\psi_m(r_2)\psi_n(r_1)\bigr]$$

For $r_1\to r_2$, which is the region where short range correlations are important, one may write $r_1=r+w/2,\;r_2=r-w/2$ and $C=C(w)$; for small $w$ one has $\Psi=\psi_m(r)\psi_n(r)\sqrt{2}+{\cal O}(w^2)$ and, as a first approximation:

\begin{eqnarray}
\int\rho^{(C,2)}(r_1,r_2)dr_2&=&(A-1)\rho^{(1)}(r_1)+2
\int\rho^{(2)}(r_1,r_2)C(r_1,r_2)dr_2\cr
&\approx& (A-1)\rho^{(1)}(r_1)+2\bigl[\rho^{(1)}(r_1)\bigr]^2\int C(w)dw
\end{eqnarray}

 \par
One can argue similarly for the three body wave function. The space components are

$$\Psi_A=\frac{1}{\sqrt{6}}Det\bigl[\psi_m(r_1)\psi_n(r_2)\psi_l(r_3)\bigr]$$

which is completely antisymmetric and corresponds to the spin quadruplet (wholly symmetric), and

\begin{eqnarray}
\Psi_1&=&\psi_l(r_3)\frac{1}{\sqrt{2}}\bigl[\psi_m(r_1)\psi_n(r_2)-\psi_m(r_2)\psi_n(r_1)\bigr]\cr
\Psi_2&=&\psi_l(r_1)\frac{1}{\sqrt{2}}\bigl[\psi_m(r_2)\psi_n(r_3)-\psi_m(r_3)\psi_n(r_2)\bigr]
\end{eqnarray}

which have mixed symmetry and correspond to the two possible spin doublets (the third possible option is a linear combination of the two above). The correlation term is not of great importance for the completely antisymmetric case, which, as a first approximation, may not need to be corrected. For
$\Psi_1,\;\Psi_2$ one introduces the correlation terms $C$ in non antisymmetric products. For example

$$\Psi_1^C=\psi_l(r_3)\frac{1}{\sqrt{2}}\bigl[\psi_m(r_1)\psi_n(r_2)-\psi_m(r_2)\psi_n(r_1)\bigr]
\bigl[1-C(u_1,u_3)-C(u_3,u_2)\bigr]$$

Unless an explicit three body correlation term is introduced, the correlation is thus again of the same kind of the two body wave function case.

An interesting possibility is that of a totally symmetric space wave function, where antisymmetry is due to the spin-isospin variables. In such a case the space wave function is

$$\frac{1}{\sqrt{6}}\sum_P\psi_l(r_1)\psi_m(r_2)\psi_n(r_3)$$

 \par\noindent
and, correspondingly, the density is

\begin{eqnarray}
\rho^{(3)}(r_1,r_2,r_3)&=&\frac{1}{6}\Big\{\sum_l|\psi_l(r_1)|^2\sum_m|\psi_m(r_2)|^2\sum_n|\psi_n(r_3)|^2\cr
&+&
\sum_l|\psi_l(r_1)|^2|\Delta(r_2,r_3)|^2+\sum_l|\psi_l(r_2)|^2|\Delta(r_3,r_1)|^2+\sum_l|\psi_l(r_3)|^2|\Delta(r_1,r_2)|^2\cr
&+&2\Re[\Delta(r_1,r_2)\Delta(r_2,r_3)\Delta(r_3,r_1)]\Big\}
\end{eqnarray}

After introducing correlations, at the first order, one has

$$\rho^{(C,3)}(r_1,r_2,r_3)=\rho^{(3)}(r_1,r_2,r_3)\bigl[1-2C(r_1,r_2)-
2C(r_2,r_3)-2C(r_3,r_1)\bigr]\;$$

As in the previous case one may introduce $r_1+r_2+r_3=3r$ and $r_1=r+w, r_2=r+w', r_3=r+w"$, where\; $w+w'+w"=0$. By expanding $\psi$ near $r$, after summing over all permutations one obtains

$$\psi\approx \psi(r)+(w+w'+w")\cdot\partial\psi/\partial r+{\cal O}(w^2)$$

and for small $w$ one may thus write

$$\Psi=\psi_l(r)\psi_m(r)\psi_n(r)\sqrt{6}+{\cal O}(w^2)$$

which allows to treat also this case on the same footing of the previous ones.

\end{document}